*UDC 535.71*

# Continual Model of Medium I: an Algorithm for the Formation of a Smooth Molecular Surface

O. Yu. Kupervasser[a,*] and N. E. Wanner[b]

[a]TRANSIST VIDEO Ltd., Skolkovo resident

[b]All-Russia Research Institute of Veterinary Sanitary, Hygiene, and Ecology, Russian Academy of Agricultural Sciences, Moscow

*e-mail: olegkup@yahoo.com

**Abstract—**In this paper, we represent a full and exhaustive algorithm for the formation of a smooth molecular solvent excluded surface (SES) and a solvent accessible surface (SAS). These surfaces are the boundaries between a molecule and a solvent. The algorithm is based on the primary and secondary rolling of molecules. The originality of this study consists in developing a *full and improved* secondary rolling algorithm, which allows us to form the optimally smooth SES of any molecule or set of molecules via the rolling of any irregularities or similar-to-irregularity situations that appear after primary rolling. For this purpose, we use the adaptive critical distance that characterizes the maximum admissible irregularity of a surface. The main problem, which is planned to be solved with the obtained surface and considered in our further papers, consists in calculating the solvation energy and its gradients for continual solvent models. This surface can also be used for demonstration purposes in molecular editors.

## 1. Introduction

The objective of our paper is to give the full and exhaustive description of an algorithm that allows us to form an optimally smooth molecular surface via primary and secondary rolling for its further use in (a) molecular editors for demonstration purposes or in (b) the calculations of the solvation energy of a molecule (the difference between its free energies in solution and vacuum) and the analytical gradients of this energy. We shall realize this smoothness via the primary rolling of a molecule with spheres, whose radius is equal to the size of a solvent molecule (for the outer surface of a molecule), and the secondary rolling of a molecule with a variable-radius sphere (for the inner surface of a molecule) with the elimination of all remaining irregularities.

We should emphasize the special importance of obtaining a smooth surface for these purposes. Really, in the case of a molecular editor, the smoothness of a surface is necessary for its triangulation and further comprehensive representation, which will not contain any physically meaningless irregularities dissipating our attention.

In the case of calculating the electrostatic component of the solvation energy and its analytical derivatives, the smoothness of a surface is a necessary algorithmic stability condition, as fictitious superficial irregularities accumulate fictitious charges, thus leading to instability in the operation of an algorithm and great numerical errors [35–37].

For this reason, the formation of a smooth surface is not only an interesting mathematical problem, but is also practically important. The primary and/or subsequent secondary rolling algorithm described in this paper was used as the base for developing the following software: PQMS [44], MSMS [34], Totrov and Abagyan's program [20], SIMS [21], TAGSS [38–40] and its improved version implemented as a subroutine of DISOLV [39–41]. These software and algorithms served as the base for creating the successfully operating molecular editor [40] and licensing the program for the calculation of the solvation energy and its derivatives [42].

What is the originality of our work in comparison with the other studies [16–19, 20, 21, 34, 38, 44] that analyze the primary and/or secondary rolling of a molecular surface? First, it consists in completeness: we have considered all the possible cases of irregularity and showed the possibility of smoothing for them. Second, we try to smooth them optimally. This means that we try not only to obtain smoothness, but also to avoid to an ultimately possible degree the appearance of surface domains that, although rather smooth, are nevertheless very similar to irregularities (i.e., to avoid the appearance of very narrow "channels" and surface domains with a very small curvature radius). For all we know, this problem has been solved fully and exhaustively only in our work.

However, we should note that the problem of constructing a smooth surface without any additional constraints may be solved trivially via the simple circumscription of a sphere around a molecule. What are these constraints?

First, the atomic radii are well-defined parameters, which may not be arbitrarily varied. When the given algorithm is used to image a molecule in a molecular editor, the radii of its atoms are determined by the sizes of electron shells (so-called van der Waals radii [1, 2], different sets of which are available in the literature [1–3]). When they are used to calculate the solvation energy, the initial ("rough") radii are specified as for the first case, but then refined so that the calculated solvation energies also correspond to their experimental values.

Second, if a molecule is submerged into a solution, the radius of solvent molecules is also a well-defined value, which may not be arbitrarily varied. This radius undoubtedly has some effect on the interface between a solution and a molecule itself and selected from the same above described reasoning as for the radius of atoms in a molecule.

Let us note that the above mentioned radii rather strongly restrict the possibility of varying the surface of a molecule. But they nevertheless hold enough opportunities for us to make it smooth.

Moreover, we wish to make it not simply smooth, but optimally smooth. In the given case, optimality is understood to mean that we (a) not only search for a smooth surface, but also try to reduce its curvature without loosing surface features and, in addition, (b) strive to decrease the Cartesian distance between ***non-adjacent surface domains***. These domains are close to each other spatially, but remote in the case of measuring the distance between them along the surface. To estimate the degree of smoothness, we specify ***the maximum critical distance***, at which non-adjacent surface domains are allowed to approach each other in the algorithm. If this approach distance is smaller than the critical distance, there occurs a similar-to-irregularity situation, which requires secondary rolling.

There exist the two types of surfaces surrounding a molecule [4]:

(1) SAS, a solvent accessible surface, is formed by the centers of solvent molecules tangent to a substrate molecule. The number of solvent molecules tangent to the surface of a molecule is proportional to the SAS area.

(2) SES, a solvent excluded surface. The volume occupied by a solvent lies *outside* the volume enveloped by this surface. The substrate itself lies completely *inside* this volume.

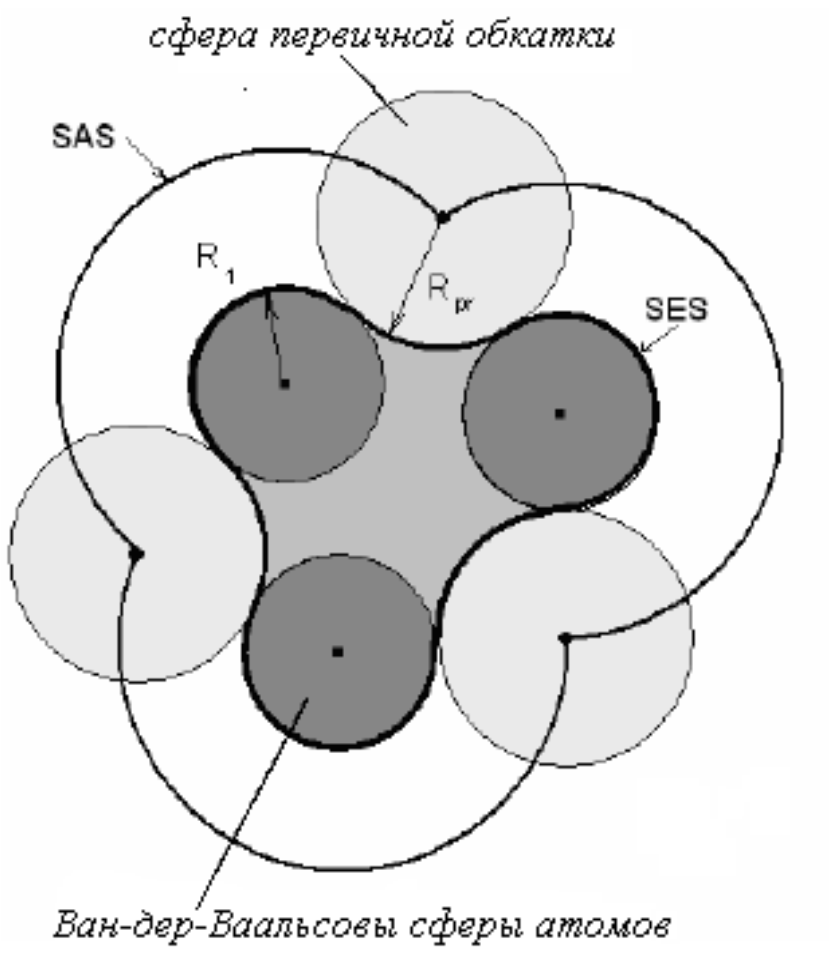

Fig.1. Primary rolling of a surface (see [41]).

Key:

сфера ван дер Вальса --> van der Waals sphere;

тороидальный сегмент первичной обкатки --> toroidal segment of primary rolling;

вогнутый сферический сегмент первичной обкатки --> concave spherical segment of primary rolling;

выпуклый сферический сегмент вторичной обкатки --> convex spherical segment of secondary rolling;

тороидальный сегмент вторичной обкатки --> toroidal segment of secondary rolling

SAS can be obtained by rolling a substrate molecule with a solvent molecule and marking the positions of its center. Rolling is the displacement of a solvent molecule along the surface of a substrate and its sequential contact with all the accessible points of this substrate (Fig. 1). For simplicity, a solvent molecule may be replaced with a rolling sphere (a sphere circumscribed around a solvent molecule) [5–6].

A molecular SES may be described as follows [7]:

(1) Smoothedly, replacing it with simple structures like [8–15]

(a) A sphere,

(b) An ellipsoid, or

(c) A cylinder;

(2) In details, reproducing all the inflections on the surface of a molecule

(a) By coating atoms with van der Waals spheres;

(b) By coating chemical groups of atom with spheres;

(c) As described in the two previous methods, but filling the remaining empty space inside SES with fictitious spheres (as implemented in the GEPOL software) [16–19]; and

(d) With the molecular electron density level surface [22] determined via quantum mechanics or the other types of functions employed for the construction of level surfaces [23–27]. This method encounters serious difficulties: a level surface may also be non-smooth; its implicit definition complicates triangulation; it is difficult to fit functions giving us a surface that is close to real and determined by the van der Waals radii of atoms; and

(e) By bridging the spheres described in (a) and (b) with convex and concave surface elements [20–21].

The smoothest and most realistic surface can be obtained by method 2e considered in our work. This method allows us to obtain SES in the same manner as SAS by rolling the outer surface of a molecule with a sphere and taking (a) the positions of points of contact between a rolling sphere and atoms, (b) the rolling sphere's geodesic arc segments that pass through two points of contact between a rolling sphere and atoms, or (c) the segments of the lower part of a rolling sphere between its geodesic arc segments that pass through two points of contact between a rolling sphere and atoms (***primary rolling***, Fig. 1). Such a technique of determining SES was first proposed in [28]. The proposed method was further developed in [20, 29–33]. However, thus obtained SES may prove to be non-smooth [34]. For its further smoothing, the inner surface of a molecule may be rolled again (***secondary rolling***, Figs. 2 and 3), as was originally proposed in [21].

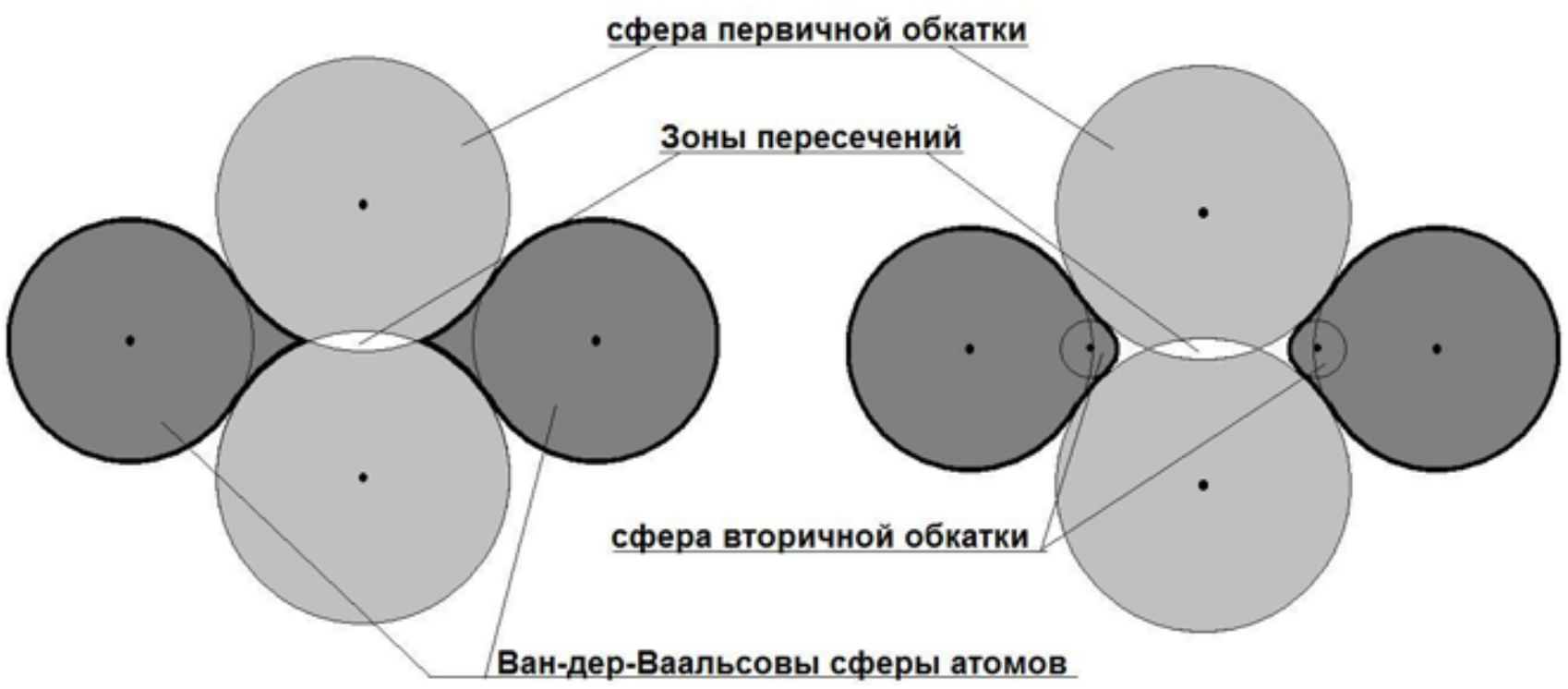

Fig. 2. Secondary rolling of a surface (see [41]).

Key:

сфера первичной обкатки --> primary rolling sphere;

Зоны пересечений --> Overlapping zones;

сфера вторичной обкатки --> secondary rolling sphere;

Ван-дер-Ваальсовы сферы атомов --> Van der Waals atomic spheres

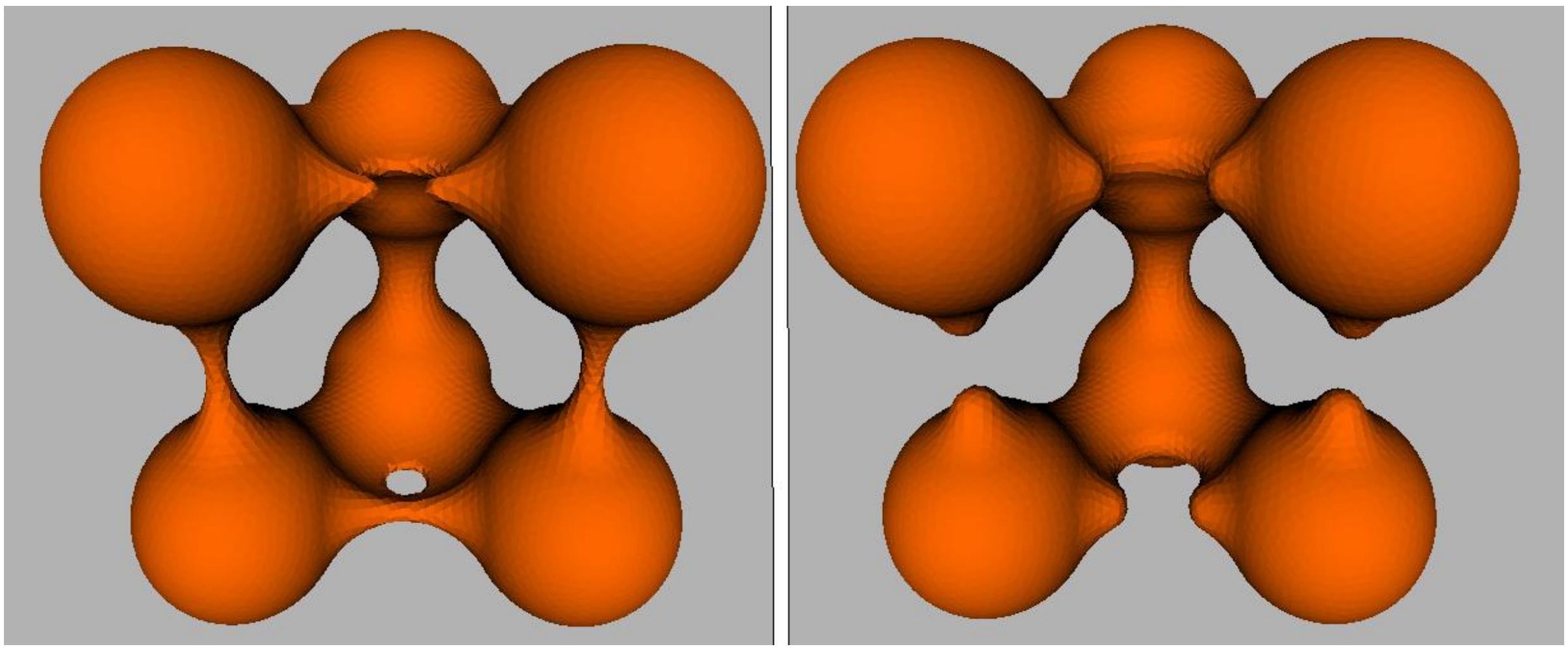

Fig. 3. Applying the method of secondary rolling to the geometric configuration of several atoms.

This may raise the question: why are we sure that secondary rolling will successfully smooth all irregularities and no tertiary, quaternary, and so forth rolling will be required? There is the following argument for this: we have no possibility of varying the radii of atoms and primary rolling spheres. They are constant predetermined values. At the same time, the secondary rolling radius is an arbitrarily selected value and may be selected so as to be various even for the rolling of different domains of the same surface. It is intuitively clear that we can "smooth" anything, selecting the secondary rolling radius as equal to an infinitely small value. Really, there are no problems for the smooth rolling of any irregularity within an infinitely small radius. However, we strive to obtain a surface that is not merely smooth, but optimally smooth. In other words, it must not contain any elements that, although rather smooth, are nevertheless similar to irregularities. For this reason, we should try to perform secondary rolling with a maximum possible radius, taking into account the geometric hindrances of a surface. This makes the algorithm of secondary rolling nontrivial enough.

A rolling surface is composed by surface segments of the two types: spherical and toroidal. They are divided into fragments of the five types: spherical elements of van der Waals atomic spheres, concave spherical elements of primary rolling, toroidal elements of primary and secondary rolling, and convex spherical elements of secondary rolling (Fig. 4).

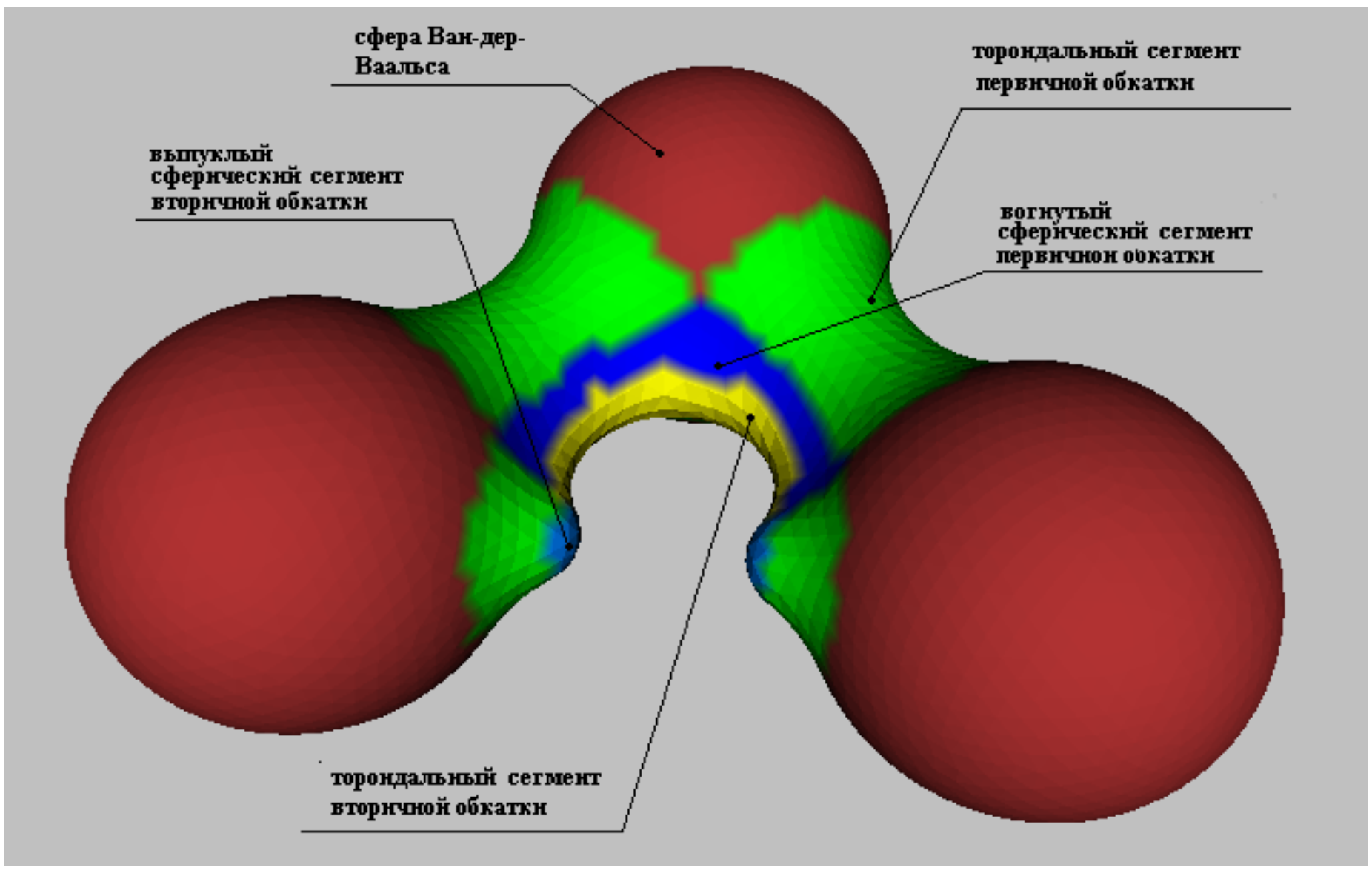

Fig. 4. Molecular surface composed of the five types of fragments: spherical elements of van der Waals atomic spheres, concave spherical elements of primary rolling, toroidal fragments of primary and secondary rolling, and convex spherical elements of secondary rolling.

Key:

сфера ван дер Вальса --> van der Waals sphere;

тороидальный сегмент первичной обкатки --> toroidal segment of primary rolling;

вогнутый сферический сегмент первичной обкатки --> concave spherical segment of primary rolling;

The primary and secondary rolling radii and the critical distance have a clear physical meaning. The primary rolling radius is equal to the radius of a sphere circumscribed around a solvent molecule. The secondary rolling radius and the above defined critical distance determine the minimum curvature of a molecular boundary. The lower limits for the secondary rolling radius and the above defined maximum critical distance are associated with the "smearing" of electron charge clouds, which can not give very acute angles and narrow necks and channels that may appear after primary rolling owing to the Heisenberg's uncertainty relation between the coordinate and impulse of an electron.

In [21] (where primary rolling is first proposed), the rolling of *some types of irregularities* that may appear after primary rolling *remain unconsidered*, namely, (a) primary rolling irregularities that are "coupled" with each other and (b) similar-to-irregularity situations, and, moreover, (c) the problem of irregularities produced by secondary rolling is not solved.

These disadvantages have been overcome in the algorithm implemented in the TAGSS (Triangulate Area Grid of Smooth Surface) software [38–40], namely, (a) a new type of irregularities considered in that algorithm is the overlapping of three primary spheres (triple secondary point), (b) *great (competitive) groups* of irregularities rolled together are considered, and (c) secondary rolling is also worth to be done for the surface domains that are regular, but similar to irregularities by their properties. These surface domains are characterized by small Cartesian spatial distances (smaller than a certain *previously specified parameter*, e.g., the critical distance) between their remote non-adjacent surface subdomains (i.e., the surface subdomains spaced at a Cartesian spatial distance, which is much longer that the distance along the surface) and have narrow channels or necks, and (d) irregularities may also appear after secondary rolling. Nevertheless, these irregularities can be eliminated merely by reducing the radius of a secondary rolling sphere until they disappear.

However, the above described methods also have some problems: (i) when the diameter of a secondary rolling sphere becomes smaller than the critical distance, it is necessary to take the above defined critical distance as equal to zero and roll only true irregularities, varying the secondary rolling radius until attaining a required value. The latter results in the appearance of domains with a very considerable curvature; (ii) formed *very large (competitive) groups of irregularities* lead to a very long operation time of the algorithm and worsen the smoothness of a surface, and there exists an increased probability of the situation that requires us to reduce the secondary rolling radius; and (iii) all the cases of several cavities (namely, the surfaces corresponding to the atomic groups lying inside large cavities or the case of several molecules) have not completely been considered.

Further steps in overcoming these problems were made in [41–43]. Correspondingly, (i) the *adaptive variable* critical distance that can gradually be reduced depending on the type of a surface was introduced; (ii) the correct and optimal mechanism of the formation of *small (competitive) groups of irregularities* was formulated, thus providing some improvement of the smoothness of a surface and the speed of secondary rolling; and (iii) the case of several molecules or groups of atoms inserted into molecular cavities was considered.

The section of our algorithm that deals with secondary rolling uses the above discussed advances [41–43] and also describes and develops them in more details. Such a secondary rolling optimally smooths all the possible irregularities and similar-to-irregularity situations that appear after primary rolling and also divides surface atoms and segments into groups corresponding to different molecular surfaces classified as belonging to diverse molecules or molecular cavity inclusions.

The methods of primary and/or secondary rolling were used as a base for developing the following software: PQMS [44], MSMS [34], Totrov and Abagyan's program [20], SIMS [21], TAGSS [38–40] and its improved version [39–41] implemented as a subroutine of DISOLV [41–43]. There also exist more complicated methods for the formation of smooth molecular surfaces in comparison with the mechanism of secondary rolling [22–27, 45–46].

## 2. Stages of the Formation of Molecular Surfaces

### 2.1. Reception of Initial Data and Filling of Inner Structures. Input and Output Data

Let there be a system consisting of several molecules submerged into a solvent. We know the coordinates and types of atoms incorporated into the system and their van der Waals radii. The first stage of the algorithm consists in determining the parameters of rolling (primary and secondary). The estimation of the rolling parameters allows us to formally describe a molecular surface as the sets of the coordinates of the positions and orientations of spherical and toroidal fragments and their geometric connectedness with each other. The simply connected surfaces of molecules incorporated into the system are initially formed. Further, the parameters of SES and SAS are determined. The arrays of atomic coordinates and van der Waals radii are the input data of the algorithm. The ***radius of a primary rolling sphere*** and the maximum ***radius of a secondary rolling sphere*** are important parameters. The ***maximum critical distance*** is also specified. It is the <u>Cartesian</u> distance, at which ***non-adjacent surface domains*** are allowed to approach each other in the algorithm. Non-adjacent surface domains are remote from each other, when the distance between them is measured <u>along</u> the surface. If their <u>Cartesian</u> approach distance is smaller than the critical distance, there occurs a similar-to-irregularity situation that requires secondary rolling.

The primary rolling radius is determined by the radius of a sphere circumscribed around a solvent molecule. The maximum secondary rolling radius is specified by a user. It must be less than half a primary rolling radius and half a van der Waals radius of each atom. The maximum critical distance must be less than the maximum secondary rolling diameter. These maximum values are adaptive and may be reduced by the algorithm itself.

The output data are the arrays of the parameters describing toroidal and spherical segments and surface atoms. The estimation of the rolling parameters allows us to formally describe a smooth molecular surface as the sets of the coordinates of the positions and orientations of spherical and toroidal fragments and their geometric connectedness with each other.

## 2.2. Determination of the Eigenbasis of a Molecule and Transformation into This Basis

For the simplicity of our consideration, let us perform some transformations into the eigenbasis of a molecule. The center of this basis is located in the center of inertia of a molecule. Its axes coincide with the principal axes of inertia of a molecule. In this calculation, all the atoms are specified to have the same "weight".

### Radius Vector of Inertia Center

$$\boldsymbol{r}_c = \frac{1}{N}\sum_{i=1}^{N}(\boldsymbol{r}_{old})_i \,, \qquad (1)$$

where $N$ is the number of atoms in a molecule and $(\boldsymbol{r}_{old})_i$ is the vector of the $i^{\text{th}}$ atom.

The tensor determining the inertia moments is calculated as

$$F = \frac{1}{N}\sum_{i-1}^{N}\begin{Vmatrix} (r_{old})_{i,x}^2 & (r_{old})_{i,x}(r_{old})_{i,y} & (r_{old})_{i,x}(r_{old})_{i,z} \\ (r_{old})_{i,y}(r_{old})_{i,x} & (r_{old})_{i,y}^2 & (r_{old})_{i,y}(r_{old})_{i,z} \\ (r_{old})_{i,z}(r_{old})_{i,x} & (r_{old})_{i,z}(r_{old})_{i,y} & (r_{old})_{i,z}^2 \end{Vmatrix}. \qquad (2)$$

Diagonalizing the matrix $F$, we obtain the diagonal matrix

$$S = G^T F G\,. \qquad (3)$$

The transformation into eigenbasis is given by the formula

$$\boldsymbol{r}_i = G((\boldsymbol{r}_{old})_i - \boldsymbol{p}_c) \text{ .} \quad (4)$$

The inverse transformation is given by the equation

$$\boldsymbol{r}_{old} = G^T \boldsymbol{r} + \boldsymbol{p}_c \text{ .} \quad (5)$$

## 2.3. Partitioning of the Outer and Inner Space of a Molecule into Overlapping Cubic Domains

The partitioning of the outer and inner space of a molecule into cubic domains represents the process of writing the indices of the atoms and surface elements of a molecule into the set of arrays, each of which corresponds to a cubic domain of certain size and position.

The partitioning into cubic domains is performed to further simplify the procedures of searching for the index of an atom or a surface element that is nearest to a point with specified coordinates. Cubes are also constructed with the purpose of accelerating the enumeration of atoms and surface elements during the formation of a surface. It usually makes no sense to consider very remote elements or atoms. Enumeration is performed only within the constructed cubes, thus reducing the space of search.

The size of a cubic domain is a constant intristic program's parameter selected from the known range of possible van der Waals atomic radii. The position and orientation of cubic domains are aligned to the eigencoordinate system's axes. We should note that cubic domains overlap each other. This allows us to find a cubic domain for any point or segment so that this point or segment lies near the center of this domain.

## 2.4. Primary and Secondary Rolling

The estimation of the rolling parameters allows us to formally describe a smooth molecular surface as the sets of the coordinates of the positions and orientations of spherical and toroidal fragments and their geometric connectedness with each other.

### 2.4.1. Primary Rolling

Primary rolling is obtained by rolling a primary rolling probe ball that simulates solvent molecules over the group of atoms forming a molecule [28]. It serves for the formation of molecular surfaces of the two types: SAS and SES. SAS is described by the positions of the center of a probe ball that simulate solvent molecules (the algorithm of its formation from SES will be considered below in part **2.7**), and SES consists of the lower fragments of a primary rolling probe ball [28] and results from the rolling over a molecule. These fragments are (a) the positions of the points of contact with atoms, (b) the rolling sphere's geodesic arc segments that pass through two points of contact between a rolling sphere and atoms, and (c) the segments of the lower part of a rolling sphere between the rolling sphere geodesic arcs' segments that pass through two points of contact between a rolling sphere and atoms (Fig. 1).

A primary rolling SES consists of spherical and toroidal fragments of the three types (Fig. 4):

(1) Convex ***van der Waals atomic sphere fragments***, when a primary rolling ball is tangent to only a single atom;

(2) Concave toroidal fragments formed by rotating a primary rolling ball around these atoms, when a primary rolling ball comes in contact only with two atoms. We shall call them ***complete primary tori***, if a rolling sphere that can pass the entire circle without thrusting against any other atoms and, otherwise, ***primary axes***.

(3) Concave spherical fragments formed by the surface of a primary rolling ball that is motionlessly fixed by atoms (more exactly, the segments of the lower part of a rolling sphere between the rolling sphere's geodesic arc segments that pass through the points of contact between a rolling sphere and atoms), when a primary rolling ball comes in contact with three atoms (or even with a greater number of atoms in singular cases, e.g., for aromatic rings). For the most typical non-singular case of three supporting atoms, such a fragment is a concave spherical triangle. For short, we shall call them ***primary triple points***.

The algorithm of primary rolling consists of the following steps:

*Step 1:* **Big primary competitive groups** are formed as follows. We select any atom and find all the neighboring atoms that are located at a distance of less than the primary rolling diameter from it, thereupon we find all the similar neighbors of these atoms, etc. The set of all such atoms forms a big primary competitive group. Then we take any atom that has not been incorporated into this group to form the next group and so on until all the atoms will be exhausted. Big primary competitive groups are formed by the sets of molecules that are not connected with each other. All the big primary competitive groups are rolled independently of each other as follows:

*Step 2:* Big primary competitive groups consisting of a single isolated atom are found. Such groups admit complete rolling without any obstacles.

*Step 3:*All the ***complete primary rolling tori*** are found for a current big primary competitive group consisting of more than one atom as follows:

- We find all the big group atom pairs, which can completely be rolled over with a primary rolling ball without thrusting against other atoms;
- A torus is divided into two independent halves, if the width of its narrowest area is smaller than the above defined maximum critical distance or this torus is self-overlapping. This is due to that such a torus will split into two unconnected pieces in the case of its self-overlapping or further secondary rolling;
- For each element (atom, complete torus or its half), we memorize its neighboring elements (atom, complete torus or its half);

*Step 4:* All the primary triple points of a big group are found as follows:

- All the triplets of atoms spaced at a distance that is less than the diameter of a primary rolling ball are enumerated;
- For each triplet of atoms, there exist either two or no any triple points;
- If any triple points exist, we check whether the primary rolling sphere forming these points overlap any other atoms. The triple points that overlap other atoms are rejected. In the case of simple contact with other atoms, a triple point is singular;
- In this manner, we enumerate all the possible triplets of atoms;
- For each element (atom, primary triple point, complete primary torus or its half), we memorize its neighboring elements (atom, primary triple point, complete primary torus or its half);

*Step 5:* The atoms that have not been involved into the previous groups are excluded from further consideration, as they are located inside the volume of a molecule and do not produce any effect on rolling. The remaining atoms will be called surface atoms.

*Step 6:*We find ***small primary competitive groups*** incorporated into big groups as follow:

- We take any above found primary triple point;
- If this triple point is not singular, all its three supporting atoms are bridged by the tori formed via the complete rotation of a rolling sphere around them. However, such a complete rolling may be hampered by other atoms. The surface of each torus may contain several unconnected segments, during the rolling of which a primary probe ball thrusts against two different atoms. Such independent segments were called above *primary axes*. They are bounded by the two utmost positions, in which a primary rolling sphere thrusts against other atoms.
- We select any atom, which is supporting for the initial triple point.
- For this supporting atom, we find all triple points and primary axes formed via its complete circular rolling with a primary ball, which is initially located at this initial triple point, until this ball encounters any obstacles. We include them in a small group.
- It should be noted that a torus, intersecting with an atom, cuts some round segments from it. After all the tori complete this "cutting", the remaining non-cut domain represents one or several unconnected polygonal spherical segments. Each similar atomic segment will also be considered as an independent element. We shall incorporate such an atomic segment into the small competitive group that is determined by the triple points and axes supported by this atomic segment.
- If this atomic segment is supporting for half a complete torus, it is also incorporated into this small group;
- We continue taking the supporting atoms of the triple points of this small group until they will completely be exhausted. The formation of the first small competitive group is thus finished.

- Then we take any triple point that has not been incorporated into the already formed small group to begin the formation of a new small group and so on until all the free triple points will be exhausted.
- The remaining free atoms (together with torus halves supported by these atoms, if any) also form independent small groups. The formation of small competitive groups is thus finished.
- Further we enumerate all the atom pairs bridged with a complete torus. If both atom segments, which are supporting for a complete torus, belong to the same small group, we incorporate this torus into this small group. If they belong to different small groups, we unite these two small groups into a single group, into which we incorporate this complete torus.
- For each element (atom, triple point, complete torus or its half, spherical and toroidal segments), we memorize its neighboring elements.

*Step 7:* Filtering of "parasitic" surfaces. The obtained small primary competitive groups incorporate one or several independent closed surfaces formed in primary rolling, namely:

(1) The outer surface of a molecule;

(2) The inner cavity, from which a primary rolling ball can not come out after entering inside it. Inner cavities usually have normals oriented inward them. Consequently, after writing the equation for the calculation of the volume of a cavity $V = \frac{1}{3}\oint_S (\boldsymbol{r} \cdot \boldsymbol{n}) dS$, we can see that it will be negative in this case. This allows us to filter away such cavities;

(3) The outer surfaces of a group of atoms "locked" inside an inner cavity. This surface may be filtered away by the following features: its inner atoms also are inner for the outer surface of a molecule, and the average radius of the outer surface is greater;

(4) The case when groups of atoms may split into several unconnected molecules (its processing is considered in part **2.5**). This takes place when primary axes are broken in the process of their self-overlapping or secondary rolling. For the complete processing of such situations, it is necessary to perform their secondary rolling first.

The procedure of primary rolling may sometimes result in the undesirable self-overlapping of a surface and kinks of SES. These irregularities may be formed by several elements of the same or various types simultaneously and classified into the two types:

(1) The self-overlapping of a primary rolling torus;

(2) The overlapping of concave secondary rolling fragments (a kink).

This may generate situations that, although smooth, are nevertheless similar to these irregular surface elements. This occurs when

(1) The width of a torus in its narrowest area is smaller than a certain empirically preselected critical distance; and

(2) The distance between concave spherical elements is smaller than this critical distance.

To overcome these problems, the method of secondary rolling was applied [21, 38–43].

### 2.4.2. Secondary Rolling

We select a secondary rolling probe ball with a diameter that is less than half a primary rolling radius and half a radius of all atoms. Using this ball, we roll the *inner* molecular surface, which was obtained after the primary rolling, only over the domains where some irregularities or similar-to-irregularity surface elements have appeared.

But, when rolling these domains, we also unavoidably roll all the primary rolling fragments connected with them. Thus, for example, when rolling the overlapping of concave spherical elements of primary rolling, we also unavoidably roll the primary rolling torus that contact with them. The other possible situations will be described below. In this way, the sets of primary rolling fragments that need secondary rolling and are united into *secondary rolling competitive groups* are formed.

The rolling of such competitive groups may produce new irregularities (or similar-to-irregularity situations) associated with the overlapping of secondary rolling ball fragments or their approach to each other at a distance that is less than the critical distance. However, there is already no need for tertiary rolling! We can eliminate these new irregularities (or similar-to-irregularity situations) via the gradual reduction of the secondary rolling radius down to the value, at which these problems disappear. There may also be the need to reduce the critical distance characterizing similar-to-irregularity situations. Such a necessity arises when the diameter of a secondary rolling ball is reduced down to a value less than this critical distance (to eliminate new irregularities or similar-to-irregularity situations). The diameter of a secondary rolling ball must always remain larger than this critical distance.

The secondary rolling SES that smooths irregularity consists of spherical and toroidal fragments of the three types (Fig. 4):

(1) When a secondary rolling ball comes in contact only with two concave spherical segments of primary rolling, the convex toroidal secondary rolling fragments that are formed via the rotation of a secondary rolling ball around these spherical segments are secondary tori. Their segments are called secondary axes.

(2) When a primary rolling ball comes in contact with three (and more in singular cases) spherical elements of primary rolling, the convex spherical fragments that are formed by the surface of a secondary rolling ball fixed motionlessly by these primary spheres are secondary triple points.

(3) When the self-overlapping of a primary rolling torus or the similar situation of a narrow torus neck occur, the pair of steady-state positions of a secondary rolling ball appears near the pair of self-overlapping points or a narrow torus neck. The secondary rolling ball's convex spherical segments that smooth irregularity form the third type—pairs of steady-state position points (Fig. 2).

We can see that the two first types of smoothing are similar to the types, which arise after primary rolling, and appear after the secondary rolling of primary spheres. The third type is new and results from the rolling of primary tori that were absent in the case of primary rolling.

The procedure for the smoothing of kinks and elimination of a self-overlapping domain via secondary rolling are illustrated in Figs. 2–4. It is performed as follows. For each earlier found small *primary* competitive group, secondary rolling is performed independently.

(1) Complete tori that have a neck width less than its critical value or are self-overlapping are rolled. A pair of steady-state position points is thus formed. This step does not produce any secondary competitive groups;

(2) Primary triple points that are overlapping or spaced at a distance less than its critical value are rolled with a secondary probe ball. If there are no any obstacles for rolling, a complete primary torus is formed and no secondary competitive group appears.

(3) In the cases different from the two previous situations, enchained irregularities appear and form secondary competitive groups. Secondary competitive groups of primary triple points are formed as follows. We take any primary triple point and find all its neighboring primary triple points at a distance less than the secondary rolling diameter from it and, thereupon, all the similar neighbors of these points, etc. The set of all similar primary triple points represents a *big secondary competitive group*. Then we take any primary triple point that has not been incorporated into this group and construct the next group and so on until all the primary triple points will be exhausted. In such *big secondary competitive groups*, we shall search for *small secondary competitive groups* of connected irregularities or similar-to-irregularity situations.

(4) Two primary triple points that overlap each other or are spaced at a distance Δ less than the critical distance are rolled with a secondary probe ball. We begin with triple points (or minimum Δ in their absence). Hereinafter, to form the next small secondary competitive groups, we shall take triple points in the ascending order of Δ. We write this current distance Δ (in the absence of enumeration) and the secondary rolling radius into the array of "reserve" critical values. Rolling produces secondary axes. A secondary probe ball may encounter some obstacles when being displaced along these axes:

(a) The displacement along this secondary axis may be hampered by another primary triple point from a big secondary competitive group. In principle, a singular case of several such primary triple points that stop rolling *simultaneously* may occur. A secondary triple point supported by three primary triple points is thus formed. New secondary axes originating from this point are also formed. We incorporate the pairs of primary triple points that are supporting for this secondary triple point and secondary axes originating from it into a small secondary competitive group for their further rolling;

(b) Primary rolling tori that belong to a big competitive group and are adjacent to this pair of primary triple points may represent obstacles for the rolling of a secondary torus. We incorporate these tori and primary axes connected with them into a small secondary competitive group for their further rolling;

(c) The rolling of such a torus produces a pair of steady-state position points. We must also roll all the pairs of primary triple points adjacent to such a torus. These pairs of primary triple points are incorporated into a small secondary competitive group for their further rolling; and

(d) We roll all the pairs of primary triple points and primary axes (both memorized earlier and newly formed by the same principle) that are incorporated into the formed small competitive group until they will completely been exhausted;

(5) During the secondary rolling described in step 4, the following situations may occur:

(a) An irregularity or a similar-to-irregularity situation that *has already been incorporated* into the other earlier formed small competitive group may fall under secondary rolling;

(b) A secondary axis that is self-overlapping or has a neck less than its critical value (width Δ) is formed;

(c) A pair of steady-state position spheres will overlap each other or approach each other at a distance (Δ) less than the critical distance; and

(d) A pair of secondary triple points will overlap each other or approach each other at a distance (Δ) less than the critical distance.

In this case, we progressively reduce the secondary rolling diameter to the maximally large diameter $D_{new}$, at which the above described situation with the rolling of an earlier rolled irregularity, an irregularity, or a similar-to-irregularity situation disappears. Then we return to step 4 and form a competitive group again. The critical distance and the secondary rolling diameter are selected as follows:

(a') Let this diameter $D_{new}$ be *larger* than a current critical distance. Then a new rolling diameter is selected so as to be equal to this distance, and the critical distance is remained unchanged. If the minimum distance Δ that has led to a problem is above zero, we memorize this distance and the found secondary rolling radius into the array of "reserve" critical distances;

(b') Let this diameter $D_{new}$ be *smaller* than a current critical distance, and the maximum "reserve" critical distance be *larger* than this new secondary rolling diameter. Then the newly selected critical distance is equal to the maximum "reserve" distance, and the secondary rolling radius is also "reserve" and corresponds to the selected critical distance. These values are excluded from "reserve";

(c') Let this diameter $D_{new}$ be *smaller* than a current critical distance, and the maximum "reserve" critical distance be smaller than this new secondary rolling diameter. Then the newly selected critical distance and secondary rolling diameter are taken so as to be equal to this secondary rolling diameter $D_{new}$; and

(d') If one of the found critical distances is very small (e.g., smaller than the value admissible by computer precision), the atomic radii are increased by 0.1 Å, and primary and secondary rolling are repeated.

6) This procedure is continued until all the small secondary competitive groups will be formed and their critical distances and secondary rolling diameters will be determined. For each element (atom, triple point, complete torus or its half, primary and secondary spherical and toroidal segments), we memorize its neighboring elements.

## 2.5. Forming the Independent Surface Atom and Segment Groups, Each of Which Corresponds to One and Only One Surface

Here we describe the algorithm for the case when the groups of atoms are splitted into several unconnected molecules (e.g., the case of docking a ligand into a substrate). In this case, the independent groups of surface atoms and segments, each of which corresponds to one and only one surface, are formed.

Small *primary* competitive groups may split into subgroups corresponding to different surfaces. This occurs when primary axes are broken in the process of their self-overlapping or by steady-state position spheres of secondary rolling.

However, such a break is unequivocal only for such a type of primary axes, whose two adjacent primary triple points (limiting its angular dimensions) lie on three or more identical supporting atoms. The other types of primary axes may remain unbroken, as they prove to be connected by the *grid* formed of one or more secondary axes after secondary rolling. We establish new connections of atoms through this newly formed grid.

Performing the analysis for connectedness (i.e., dividing the atoms of small primary competitive groups into other subgroups in view of the break of primary axes (tori) and, sometimes, restoring the connectedness of these subgroups through the grid of secondary tori), we find the primary surface atom and segment groups that will correspond to different surfaces.

## 2.6. Determining the Parameters of Spherical and Toroidal Segments

### 2.6.1. Determining the Parameters of the Convex Spherical Segments of Surface Atoms

Convex spherical fragments are the parts of van der Waals atomic spheres oriented towards a solvent. To describe these spherical fragments for each atom, we use the set of the vectors characterizing their orientation towards the neighboring contact atoms and the set of the "forbidden" angles related with each orientation. Atoms, whose spheres are spaced at a distance less than the diameter of a primary rolling sphere $2R_{pr}$, are considered to be contact. The angle, at which the radius of the arc trajectory of the center of a rolling sphere tangent to a pair of atoms—current and second contact atoms—can be seen, is called "forbidden". The part of the van der Waals sphere of a given atom outside all the "competitive cones" is the sought convex spherical fragment of a molecular surface. We should note that forbidden angles can divide an atomic sphere not into one, but also into two or even more spherical fragments $S_{R_j}$ that are not connected with each other.

Let $r_0$ be the vector of the coordinates of a current atom, and $r_i$ be the vector of the coordinate of a contact atom, $R_0$ be the radius of a current atom, and $R_i$ be the radius of a contact atom.

For each atom, the arrays of unit vectors $\vec{d}_i$ and forbidden angles $\gamma_i$ are defined. The orientation vector is determined from the condition

$$\boldsymbol{d}_i = \frac{\boldsymbol{r}_i - \boldsymbol{r}_0}{|\boldsymbol{r}_i - \boldsymbol{r}_0|}, \tag{6}$$

and the forbidden angle is found as

$$cos(\gamma_i) = \frac{b^2 - a^2 - c^2}{2ac}, \quad a = R_0 + R_{pr}, \quad b = R_i + R_{pr}, \quad c = |\boldsymbol{r}_i - \boldsymbol{r}_0|, \tag{7}$$

where the indices 0 and *i* denote current and contact atoms, respectively. The algorithm of determining the parameters of convex spherical elements consists in enumerating all the pairs of contact atoms and calculating the set of orientation vectors and corresponding forbidden angles for each atom.

### 2.6.2. Determining the Array of the Center Coordinates of Primary Concave Spherical Fragments (Primary Spherical Bridging) between the Triplets of Atoms

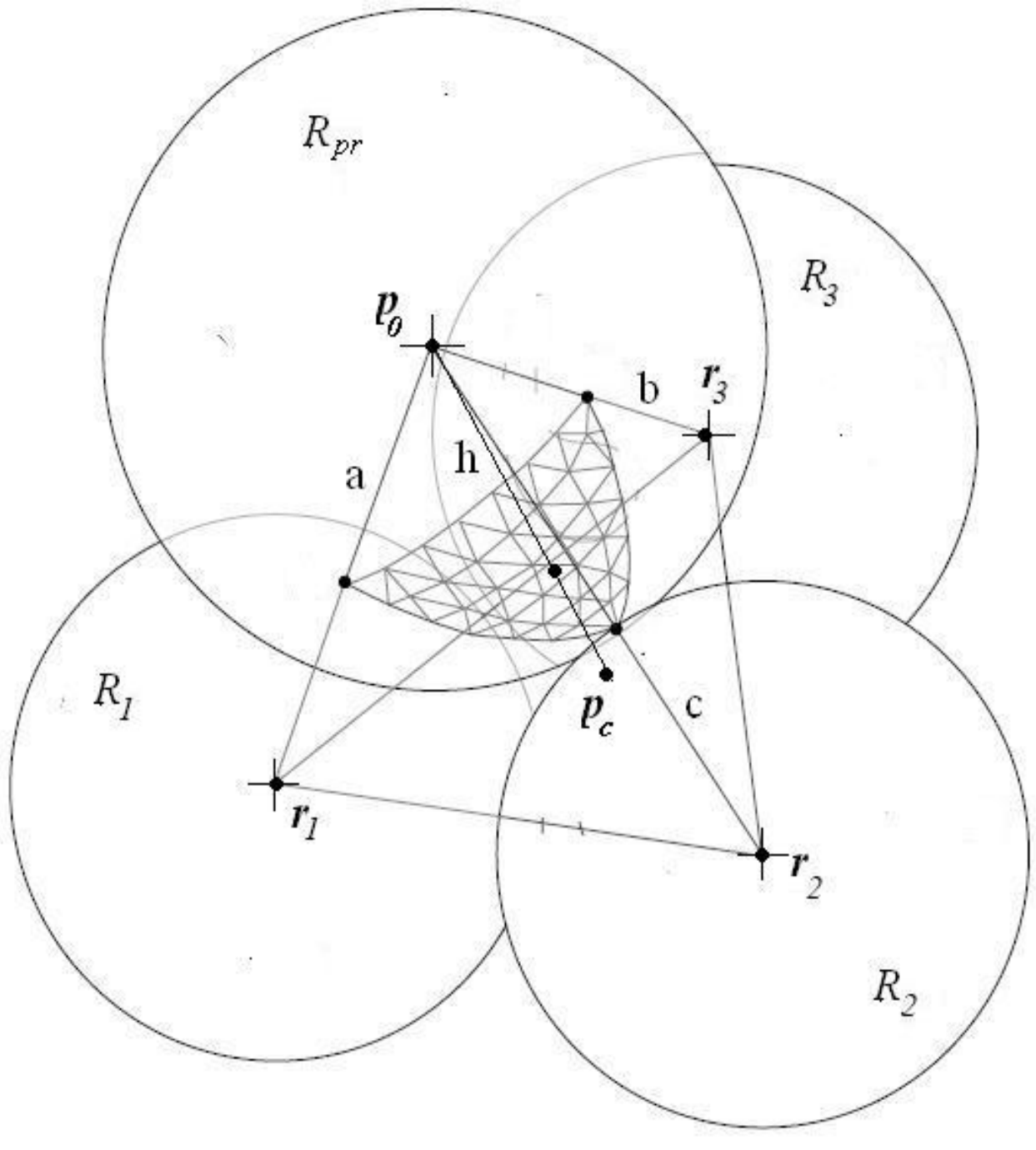

Fig. 5. Geometric interpretation of the problem of determining the center of a sphere tangent to three atoms.

The concave spherical elements of SES (Fig. 5) are shaped as a spherical triangle and formed upon the contact of a rolling sphere with three atoms simultaneously (except rare singular cases). The points of contact between a rolling sphere and atomic spheres determine the apices of a spherical triangle constituting a part of the rolling sphere.

The coordinates of the vector $\boldsymbol{p}_0$ of the center of a sphere $R_{pr}$ supported by three atoms with the coordinates $\boldsymbol{r}_1, \boldsymbol{r}_2$, and $\boldsymbol{r}_3$ are calculated by the below given formulas.

For the side faces of a pyramid,

$$a = R_1 + R_{pr};\ b = R_2 + R_{pr};\ c = R_3 + R_{pr}, \qquad (8)$$

where $h$ is the height of a pyramid and $\boldsymbol{p}_c$ is the radius vector of the base height of a pyramid.

$$\boldsymbol{z} = \frac{[(\boldsymbol{r}_3 - \boldsymbol{r}_1) \times (\boldsymbol{r}_2 - \boldsymbol{r}_1)]}{\|[(\boldsymbol{r}_2 - \boldsymbol{r}_1) \times (\boldsymbol{r}_3 - \boldsymbol{r}_1)]\|}, \qquad (9)$$

$$\boldsymbol{r} = \boldsymbol{p}_c \pm \boldsymbol{z} h, \qquad (10)$$

$$\begin{cases} (\boldsymbol{r}-\boldsymbol{r}_1)^2 = a^2 \\ (\boldsymbol{r}-\boldsymbol{r}_2)^2 = b^2 \\ (\boldsymbol{r}-\boldsymbol{r}_3)^2 = c^2 \end{cases} \quad \Leftrightarrow \tag{11}$$

$$\begin{cases} (\boldsymbol{p}_c-\boldsymbol{r}_1)^2 = a^2 - h^2 \\ (\boldsymbol{p}_c-\boldsymbol{r}_2)^2 = b^2 - h^2 \\ (\boldsymbol{p}_c-\boldsymbol{r}_3)^2 = c^2 - h^2 \end{cases} \quad => \tag{12}$$

$$\begin{cases} (\mathbf{r}_2-\mathbf{r}_1)\cdot\mathbf{p}_c = (a^2 - b^2 - \vec{r}_1^{\,2} + \vec{r}_2^{\,2})/2 \\ (\mathbf{r}_3-\mathbf{r}_1)\cdot\mathbf{p}_c = (a^2 - c^2 - \vec{r}_1^{\,2} + \vec{r}_3^{\,2})/2 \end{cases}, \tag{13}$$

$$\boldsymbol{p}_c = \boldsymbol{r}_1 + \alpha[-\mathbf{z}\times(\boldsymbol{r}_2-\boldsymbol{r}_1)] + \beta[\mathbf{z}\times(\boldsymbol{r}_3-\boldsymbol{r}_1)], \tag{14}$$

$$\boldsymbol{p}_c = \boldsymbol{r}_1 + [\mathbf{z}\times(\gamma\boldsymbol{r}_1 - \alpha\boldsymbol{r}_2 + \beta\boldsymbol{r}_3)], \tag{15}$$

$$\begin{cases} \alpha = \dfrac{(a^2 - c^2 + (\boldsymbol{r}_3-\boldsymbol{r}_1)^2)}{2\left\|[(\boldsymbol{r}_2-\boldsymbol{r}_1)\times(\boldsymbol{r}_3-\boldsymbol{r}_1)]\right\|} \\ \beta = \dfrac{(a^2 - b^2 + (\boldsymbol{r}_2-\boldsymbol{r}_1)^2)}{2\left\|[(\boldsymbol{r}_2-\boldsymbol{r}_1)\times(\boldsymbol{r}_3-\boldsymbol{r}_1)]\right\|} \\ \gamma = \alpha - \beta = \dfrac{(b^2 - c^2 - (\boldsymbol{r}_2-\boldsymbol{r}_1)^2 + (\boldsymbol{r}_3-\boldsymbol{r}_1)^2)}{2\left\|[(\boldsymbol{r}_2-\boldsymbol{r}_1)\times(\boldsymbol{r}_3-\boldsymbol{r}_1)]\right\|} \end{cases}, \tag{16}$$

$$[(\boldsymbol{r}_3-\boldsymbol{r}_1)\times(\boldsymbol{r}_2-\boldsymbol{r}_1)] = [(\boldsymbol{r}_2-\boldsymbol{r}_3)\times(\boldsymbol{r}_1-\boldsymbol{r}_3)] = [(\boldsymbol{r}_1-\boldsymbol{r}_2)\times(\boldsymbol{r}_3-\boldsymbol{r}_2)], \tag{17}$$

$$\boldsymbol{p}_c = \frac{(\boldsymbol{r}_1+\boldsymbol{r}_2+\boldsymbol{r}_3)}{3} + \frac{1}{6}\left[\begin{pmatrix} \dfrac{3(b^2-c^2)+(\boldsymbol{r}_3-\boldsymbol{r}_1)^2-(\boldsymbol{r}_2-\boldsymbol{r}_1)^2}{\left\|[(\boldsymbol{r}_3-\boldsymbol{r}_1)\times(\boldsymbol{r}_2-\boldsymbol{r}_1)]\right\|^2}[[(\boldsymbol{r}_3-\boldsymbol{r}_1)\times(\boldsymbol{r}_2-\boldsymbol{r}_1)]\times\boldsymbol{r}_1] + \\ +\dfrac{3(c^2-a^2)+(\boldsymbol{r}_1-\boldsymbol{r}_2)^2-(\boldsymbol{r}_3-\boldsymbol{r}_2)^2}{\left\|[(\boldsymbol{r}_1-\boldsymbol{r}_2)\times(\boldsymbol{r}_3-\boldsymbol{r}_2)]\right\|^2}[[(\boldsymbol{r}_1-\boldsymbol{r}_2)\times(\boldsymbol{r}_3-\boldsymbol{r}_2)]\times\boldsymbol{r}_2] + \\ +\dfrac{3(a^2-b^2)+(\boldsymbol{r}_2-\boldsymbol{r}_3)^2-(\boldsymbol{r}_1-\boldsymbol{r}_3)^2}{\left\|[(\boldsymbol{r}_2-\boldsymbol{r}_3)\times(\boldsymbol{r}_1-\boldsymbol{r}_3)]\right\|^2}[[(\boldsymbol{r}_2-\boldsymbol{r}_3)\times(\boldsymbol{r}_1-\boldsymbol{r}_3)]\times\vec{r}_3] \end{pmatrix}\right], \tag{18}$$

$$h^2 = \frac{a^2+b^2+c^2-(\boldsymbol{r}_1^2+\boldsymbol{r}_2^2+\boldsymbol{r}_3^2)+2\boldsymbol{p}_c\cdot(\boldsymbol{r}_1+\boldsymbol{r}_2+\boldsymbol{r}_3)}{3} - \boldsymbol{p}_c^2. \tag{19}$$

For each primary rolling sphere, the arrays of unit vectors $\boldsymbol{d}_i$ and forbidden angles $\gamma_i$ are defined. They are related with the secondary tori of rolling over spherical elements. The orientation vector is found from the condition

$$\boldsymbol{d}_i = \frac{\boldsymbol{r}_i - \boldsymbol{p}_0}{|\boldsymbol{r}_i - \boldsymbol{p}_0|}, \tag{20}$$

and the forbidden angle is determined as

$$cos(\gamma_i) = \frac{b'^2 - a'^2 - c'^2}{2a'c'}, \quad a' = R_{pr} + R_{sec}, \quad b' = R_{pr} + R_{sec}, \quad c' = |\boldsymbol{r}_i - \boldsymbol{p}_0|, \tag{21}$$

where the indices 0 and *i* denote current and contact primary spheres, respectively.

The algorithm of determining the parameters of concave spherical elements consist in enumerating all the pairs of concave spherical elements, for which secondary rolling is performed, and calculating the set of orientation vectors and corresponding forbidden angles for each concave spherical element.

### 2.6.3. Determining the Arrays of the Parameters of Toroidal Fragments between Atomic Pairs

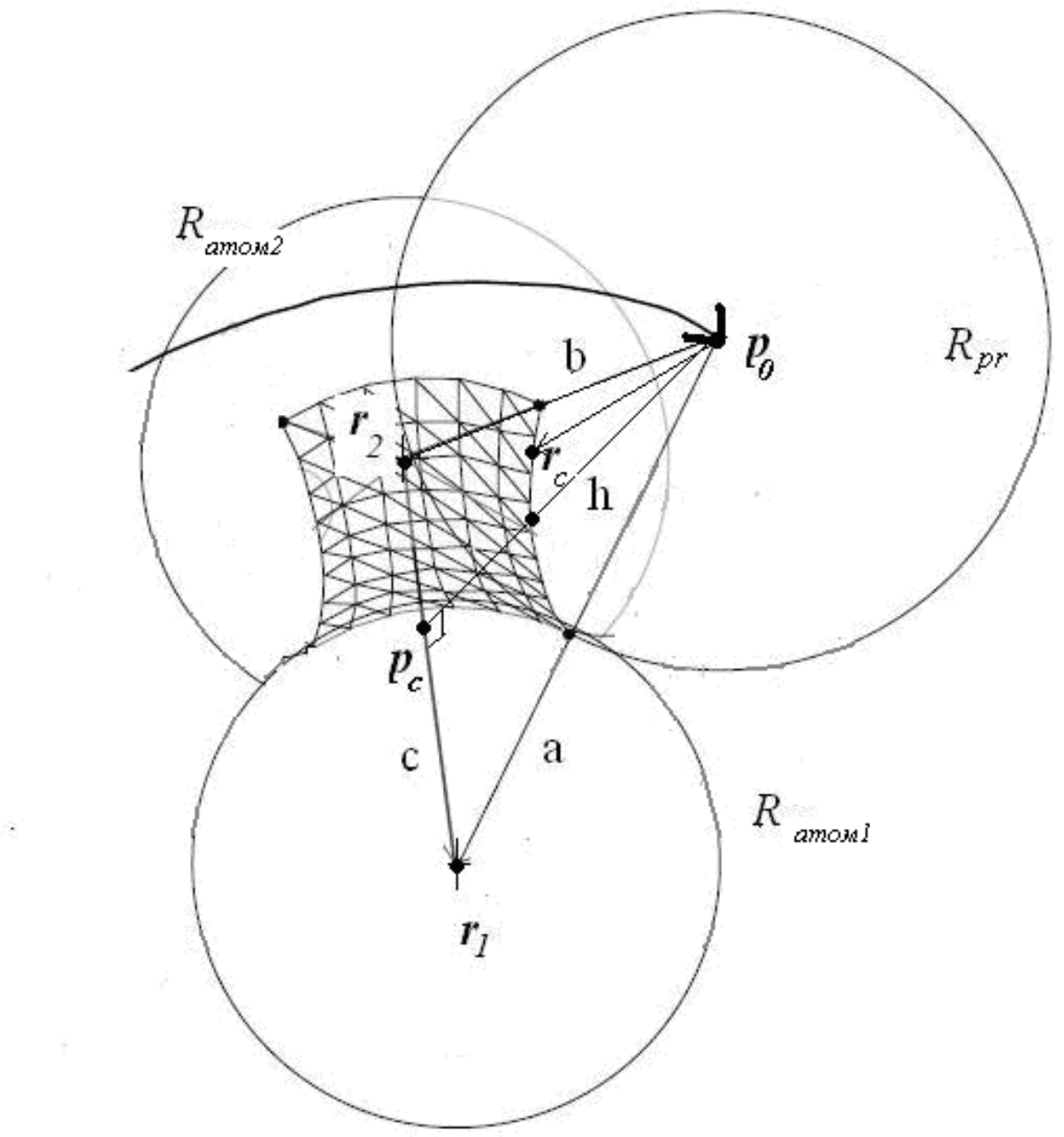

Fig. 6. Rolling of two atoms with a primary probe ball. Formation of a primary torus.

Key:

$R_{атом1}$ --> $R_{atom1}$ ;

$R_{атом2}$ --> $R_{atom2}$ ;

Primary toroidal surface fragments (Fig. 6) are formed during the rolling of two atoms with a primary sphere. In this case, the center of a rolling sphere describes a portion of circle or a complete circle depending on the location of other atoms. A toroidal fragment is confined between the points of contact between the sphere of two atoms and the positions, in which a rolling sphere thrusts against any other atom, i.e., the points of contact between a rolling sphere and three atoms simultaneously.

Toroidal fragments are described using the following parameters:

(1) The center $\boldsymbol{p}_c$ and radius $h$ of the circle, along which the center of a rolling sphere describes an arc trajectory;

(2) The local basis *x, y,* and *z*, whose *z* axis is oriented parallelly to the line connecting the centers of supporting atoms, *x* and *y* axes define the plane of the arc described by the center of a rolling sphere, and *x* axis is oriented towards one of the atoms that represent an obstacle for the rolling over the given pair of atoms;

(3) The angles $\alpha$ and $\beta$ with respect to the *x* axis for the beginning and end of the circular trajectory of the center of a rolling sphere; and

(4) The angle $\gamma$ characterizing the maximum angular deviation from the line determined by the *z* axis for the radius vector from the current point of a torus to the current center of a rolling sphere $\boldsymbol{p_0}$ - $\boldsymbol{r}_s$. At this deviation, it can not yet be cut from a torus by one of the steady-state position spheres, if the secondary rolling of a torus was required.

For the elements located on the toroidal bridge between the atoms $\boldsymbol{r}_1$ and $\boldsymbol{r}_2$, we use the radii of primary and secondary rolling spheres $R_{pr}$ and $R_{sec}$ and the radii of the first and second atoms $R_1$ and $R_2$, respectively.

The intermediate calculations are performed using the following formulas:

$$a = R_1 + R_{pr}; b = R_2 + R_{pr}; c = \left|\boldsymbol{r}_2 - \boldsymbol{r}_1\right|, \tag{22}$$

where *c* is the distance between atoms.

The radius of the circular trajectory described by the center of a rolling sphere is calculated as

$$h = \frac{1}{2c}\sqrt{4a^2c^2 - \left(a^2 + c^2 - b^2\right)^2}\,. \tag{23}$$

The center of this circular trajectory is determined by the expression

$$\boldsymbol{p}_c = \tfrac{1}{2}\left(\boldsymbol{r}_1 + \boldsymbol{r}_2\right) + \frac{\left(\boldsymbol{r}_2 - \boldsymbol{r}_1\right)\left(a^2 - b^2\right)}{2c^2}\,. \tag{24}$$

The local orthonormalized basis is found as

$$\boldsymbol{z} = \frac{\left(\boldsymbol{r}_2 - \boldsymbol{r}_1\right)}{c}, \quad (\boldsymbol{x}\cdot\boldsymbol{z}) = 0, \quad (\boldsymbol{y}\cdot\boldsymbol{z}) = 0, \quad (\boldsymbol{x}\cdot\boldsymbol{y}) = 0\,. \tag{25}$$

If the atom with the coordinates $\vec{r}_3$ bounds the rolling over the atoms $\vec{r}_1$ and $\vec{r}_2$, the basis vectors are determined via the following equations:

$$\boldsymbol{l} = \frac{\boldsymbol{r}_3 - \boldsymbol{p}_c}{|\boldsymbol{r}_3 - \boldsymbol{p}_c|}, \quad \boldsymbol{y} = [\boldsymbol{z} \times \boldsymbol{l}], \quad \boldsymbol{x} = [\boldsymbol{z} \times \boldsymbol{y}], \tag{26}$$

The angles $\alpha$ and $\beta$ are calculated as

$$\alpha = arccos(\boldsymbol{x} \cdot \frac{\boldsymbol{p}_{0,\alpha} - \boldsymbol{p}_c}{|\boldsymbol{p}_{0,\alpha} - \boldsymbol{p}_c|}), \quad \beta = arccos(\boldsymbol{x} \cdot \frac{\boldsymbol{p}_{0,\beta} - \boldsymbol{p}_c}{|\boldsymbol{p}_{0,\beta} - \boldsymbol{p}_c|}), \tag{27}$$

where $\boldsymbol{p}_{0,\alpha}$ and $\boldsymbol{p}_{0,\beta}$ are the coordinates of the center of a rolling sphere tangent to three atoms at the beginning and end of the arc of rolling over a pair of atoms. The angle $\gamma$ is calculated as

$$sin(\gamma) = \frac{h}{R_{pr} + R_{sec}}. \tag{28}$$

### 2.6.4. Determining the Array of the Center and Radius Coordinates of Secondary Spherical Fragments (Secondary Spherical Bridging) between Triplets of Primary Spherical Fragments

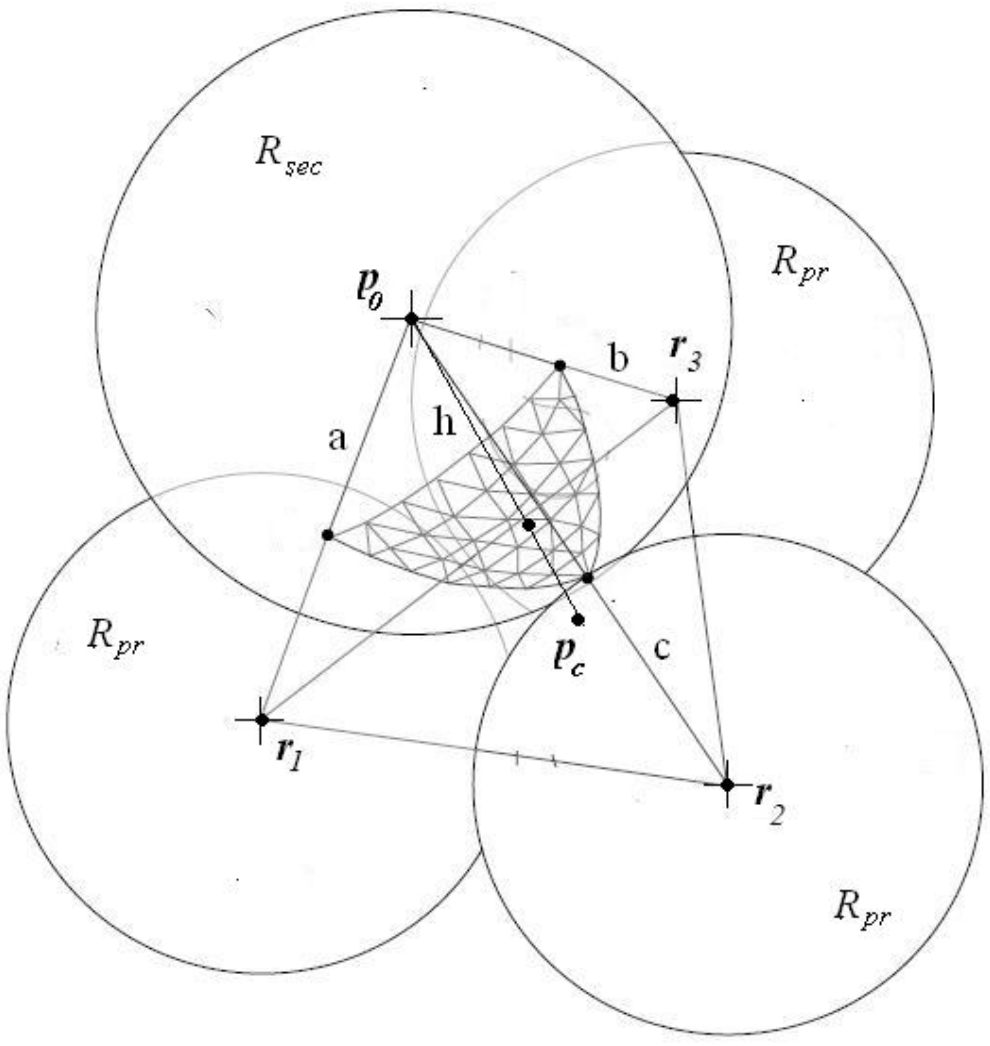

Fig. 7. Geometric interpretation of the problem of determining the center of a secondary probe ball tangent to three primary rolling spheres.

The coordinates of the vector $\boldsymbol{p}_0$ of the center of a secondary sphere supported by three primary rolling spheres with the coordinates $\boldsymbol{r}_1$, $\boldsymbol{r}_2$ and $\boldsymbol{r}_3$ are calculated by the following formulas. The radii vectors of the positions of the centers of supporting primary spheres $\boldsymbol{r}_1$, $\boldsymbol{r}_2$, and $\boldsymbol{r}_3$ determine the base of a pyramid, and the center of a secondary probe ball $\boldsymbol{r}$ determines its apex. This case is similar to primary spherical bridging. It is even simpler, as all the pyramid edges originating from its apex are equal in the current case (Fig 7).

The radius vector of the center of a probe ball may be resolved into the two components, such as the radius vector of the pyramid base height $\boldsymbol{p}_c$ and the unit vector perpendicular to the base of a pyramid $\boldsymbol{z}$, i.e.,

$$\boldsymbol{r} = \boldsymbol{p}_c \pm \boldsymbol{z} h \ . \tag{29}$$

These two radii vectors are determined by the formulas

$$\boldsymbol{z} = \frac{[(\boldsymbol{r}_3 - \boldsymbol{r}_1) \times (\boldsymbol{r}_2 - \boldsymbol{r}_1)]}{\|[(\boldsymbol{r}_2 - \boldsymbol{r}_1) \times (\boldsymbol{r}_3 - \boldsymbol{r}_1)]\|}\ , \tag{30}$$

$$\boldsymbol{p}_c = \frac{(\boldsymbol{r}_1 + \boldsymbol{r}_2 + \boldsymbol{r}_3)}{3} + \frac{1}{6}\left[\left(\begin{array}{l} \dfrac{(\boldsymbol{r}_3 - \boldsymbol{r}_1)^2 - (\boldsymbol{r}_2 - \boldsymbol{r}_1)^2}{\|[(\boldsymbol{r}_3 - \boldsymbol{r}_1) \times (\boldsymbol{r}_2 - \boldsymbol{r}_1)]\|^2}[\,[(\boldsymbol{r}_3 - \boldsymbol{r}_1) \times (\boldsymbol{r}_2 - \boldsymbol{r}_1)] \times \boldsymbol{r}_1\,] + \\ + \dfrac{(\boldsymbol{r}_1 - \boldsymbol{r}_2)^2 - (\boldsymbol{r}_3 - \boldsymbol{r}_2)^2}{\|[(\boldsymbol{r}_1 - \boldsymbol{r}_2) \times (\boldsymbol{r}_3 - \boldsymbol{r}_2)]\|^2}[\,[(\boldsymbol{r}_1 - \boldsymbol{r}_2) \times (\boldsymbol{r}_3 - \boldsymbol{r}_2)] \times \boldsymbol{r}_2\,] + \\ + \dfrac{(\boldsymbol{r}_2 - \boldsymbol{r}_3)^2 - (\boldsymbol{r}_1 - \boldsymbol{r}_3)^2}{\|[(\boldsymbol{r}_2 - \boldsymbol{r}_3) \times (\boldsymbol{r}_1 - \boldsymbol{r}_3)]\|^2}[\,[(\boldsymbol{r}_2 - \boldsymbol{r}_3) \times (\boldsymbol{r}_1 - \boldsymbol{r}_3)] \times \boldsymbol{r}_3\,] \end{array}\right)\right], \tag{31}$$

$$h^2 = (R_{pr} + R_{sec})^2 - (\boldsymbol{r}_1^2 + \boldsymbol{r}_2^2 + \boldsymbol{r}_3^2)/3 - \boldsymbol{p}_c^2 + 2\boldsymbol{p}_c \cdot \frac{(\boldsymbol{r}_1 + \boldsymbol{r}_2 + \boldsymbol{r}_3)}{3}, \tag{32}$$

where $R_{sec}$ is the radius of a secondary rolling sphere.

### 2.6.5. Determining the Array of the Center and Radius Coordinates of Pairs of Secondary Spherical Fragments (Steady-State Position Spheres) That Smooth Primary Rolling Tori in the Case of Their Self-Overlapping or Narrow Necks

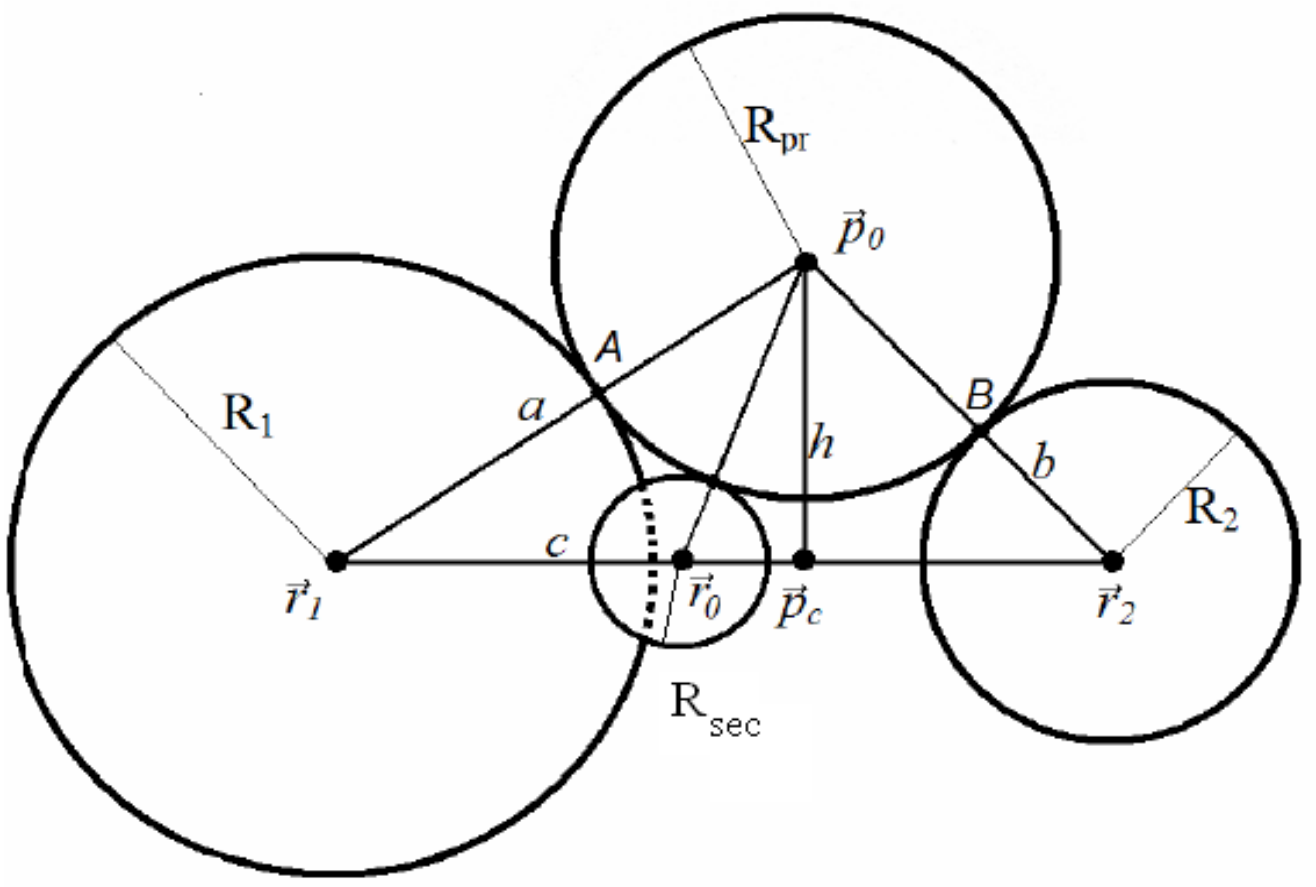

Fig. 8. Secondary steady-state position spheres (see [41]).

For the points located on the surface of steady-state position spheres (Fig. 8), the distance between the centers of a sphere and a torus $d$ is calculated as

$$d = \sqrt{(R_{pr} + R_{\sec})^2 - h^2}\ , \qquad (33)$$

and the radii vectors of the centers of steady-state position spheres $\boldsymbol{r}_0$ is determined as

$$\boldsymbol{r}_0 = \boldsymbol{p}_c \pm d\boldsymbol{z}\ . \qquad (34)$$

**2.6.6. Determining the Array of the Center and Radius Coordinates of Secondary Spherical Fragments (Secondary Spherical Bridging) between Triplets of Primary Spherical Fragments**

Secondary toroidal surface fragments are formed during the rolling of two primary spherical segments with a secondary rolling sphere (Fig. 9). In this case, the center of a secondary rolling sphere describes a portion of circle or a complete circle depending on the location of other primary spheres or tori. A secondary toroidal fragment is confined between the point of contact between a secondary sphere and two primary spheres and the positions, in which a secondary sphere thrusts against any other primary sphere (or even several spheres *simultaneously*) or primary torus.

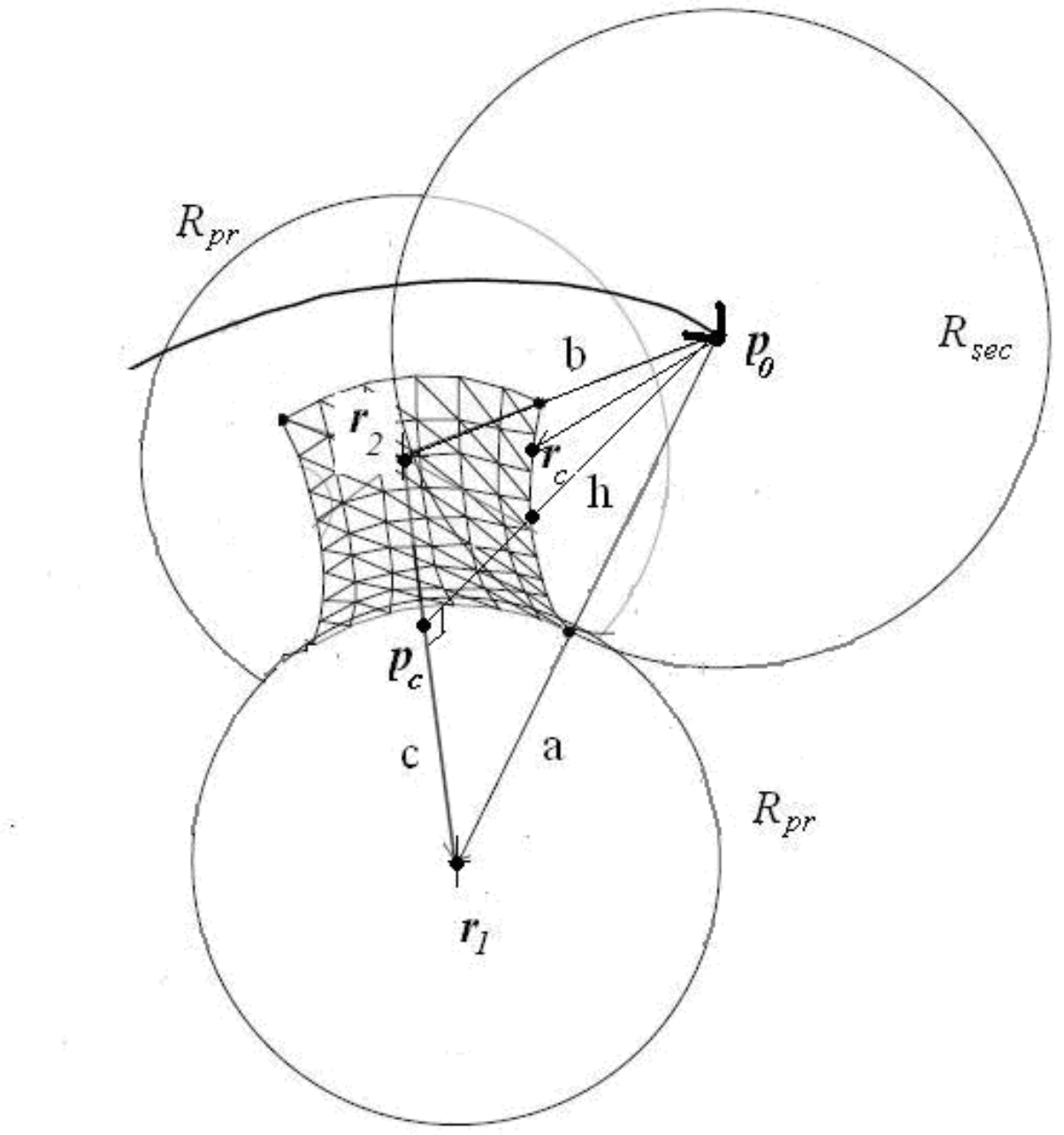

Fig. 9. Secondary rolling of two primary rolling spheres. Secondary rolling tori.

Secondary toroidal fragments are described using the following parameters:

(1) The center $\boldsymbol{p}_c$ and radius $h$ of the circle, along which the center of a secondary rolling sphere describes an arc trajectory;

(2) The local basis $x,y,z$, whose $z$ axis is oriented parallelly to the line connecting the centers of primary supporting spheres, and $x$ and $y$ axes determine the plane of the arc described by the center of a secondary rolling sphere. The $x$ axis is oriented towards either one of the primary rolling spheres that represent an obstacle for the rolling over the given pair of atoms of primary rolling spheres or the center of a secondary sphere that restricts rotation (for a primary torus that restricts displacement).

(3) The angles $\alpha$ and $\beta$ relative to the $x$ axis for the beginning and end of the arc trajectory described by the center of a rolling sphere. The plane shown in Fig. 9 is determined by the surface point $\boldsymbol{r}_s$ and the centers of primary rolling spheres $\boldsymbol{r}_1$ and $\boldsymbol{r}_2$: $R_{pr}$ and $R_{sec}$ are the radii of primary and secondary rolling spheres, respectively.

The edges of a triangle formed by the centers of spheres and their derivatives are determined as

$$a = b = R_{pr} + R_{sec}\ ;\ c = \left|\boldsymbol{r}_2 - \boldsymbol{r}_1\right| . \tag{35}$$

The height of this triangle or, equivalently, the radius of the circular trajectory described by the center of a secondary rolling sphere is calculated as

$$h = \sqrt{a^2 - \frac{c^2}{4}} \quad . \tag{36}$$

The center of the circular trajectory described by the center of a secondary rolling sphere or, equivalently, the radius vector of the triangle base height point $\boldsymbol{p}_c$ is found as

$$\boldsymbol{p}_c = \frac{\boldsymbol{r}_1 + \boldsymbol{r}_2}{2} . \tag{37}$$

The local basis of a paired bridge $\boldsymbol{x}, \boldsymbol{y}, \boldsymbol{z}$ , whose coordinate center is located at the base height point $\boldsymbol{p}_c$, and the *z* axis is directed from the first primary supporting sphere to the second primary supporting sphere) is determined as

$$\boldsymbol{z} = \frac{\boldsymbol{r}_2 - \boldsymbol{r}_1}{c}, \ \ (\boldsymbol{x} \cdot \boldsymbol{z}) = 0, \ \ (\boldsymbol{y} \cdot \boldsymbol{z}) = 0, \ \ (\boldsymbol{x} \cdot \boldsymbol{y}) = 0 . \tag{38}$$

If a primary sphere with the coordinates $\vec{r}_3$ restricts the rolling over the primary spheres $\boldsymbol{r}_1$ and $\boldsymbol{r}_2$, the basis vectors are found as

$$\boldsymbol{l} = \frac{\boldsymbol{r}_3 - \boldsymbol{p}_c}{|\boldsymbol{r}_3 - \boldsymbol{p}_c|}, \ \ \boldsymbol{y} = [\boldsymbol{z} \times \boldsymbol{l}], \ \ \boldsymbol{x} = [\boldsymbol{z} \times \boldsymbol{y}]. \tag{39}$$

If a secondary sphere restricts the rolling over the primary spheres $\boldsymbol{r}_1$ and $\boldsymbol{r}_2$ and is selected to specify the direction of the *x* axis, the basis vectors are found as

$$\boldsymbol{x} = \frac{\boldsymbol{p}_{0,\alpha} - \boldsymbol{p}_c}{|\boldsymbol{p}_{0,\alpha} - \boldsymbol{p}_c|}, \ \ \boldsymbol{y} = [\boldsymbol{z} \times \boldsymbol{x}], \tag{40}$$

where $\boldsymbol{p}_{0,\alpha}$ is the radius vector of the center of a secondary restricting sphere.

The angles $\alpha$ and $\beta$ are calculated as

$$\alpha = arccos(\boldsymbol{x} \cdot \frac{\boldsymbol{p}_{0,\alpha} - \boldsymbol{p}_c}{|\boldsymbol{p}_{0,\alpha} - \boldsymbol{p}_c|}), \ \ \beta = arccos(\boldsymbol{x} \cdot \frac{\boldsymbol{p}_{0,\beta} - \boldsymbol{p}_c}{|\boldsymbol{p}_{0,\beta} - \boldsymbol{p}_c|}), \tag{41}$$

where $\boldsymbol{p}_{0,\alpha}$ and $\boldsymbol{p}_{0,\beta}$ are the coordinates of the center of a secondary rolling sphere tangent to three primary spheres at the beginning and end of the arc of rolling around a pair of primary spheres or the coordinates of the centers of secondary steady-state position spheres.

The projection of a normal onto the local basis for secondary rolling tori (the normal is oriented inward a torus) is determined as

$$\begin{aligned} & \boldsymbol{n}_s \cdot \boldsymbol{z} = -\gamma_s , \\ & (\boldsymbol{n}_s \cdot \boldsymbol{x}) = ((-\sqrt{\boldsymbol{n}_{sx}^2 + \boldsymbol{n}_{sy}^2}\,\boldsymbol{e}_s) \cdot \boldsymbol{x}) = -\alpha_s \sqrt{1-\gamma_s^2} , \\ & (\boldsymbol{n}_s \cdot \boldsymbol{y}) = ((-\sqrt{\boldsymbol{n}_{sx}^2 + \boldsymbol{n}_{sy}^2}\,\boldsymbol{e}_s) \cdot \boldsymbol{y}) = -\beta_s \sqrt{1-\gamma_s^2} . \end{aligned} \tag{42}$$

$$\boldsymbol{n}_s = -\alpha_s \cdot \sqrt{1-\gamma_s^2} \cdot \boldsymbol{x} - \beta_s \cdot \sqrt{1-\gamma_s^2} \cdot \boldsymbol{y} - \gamma_s \cdot \boldsymbol{z}. \tag{43}$$

## 2.7. Solvent Accessible Surface

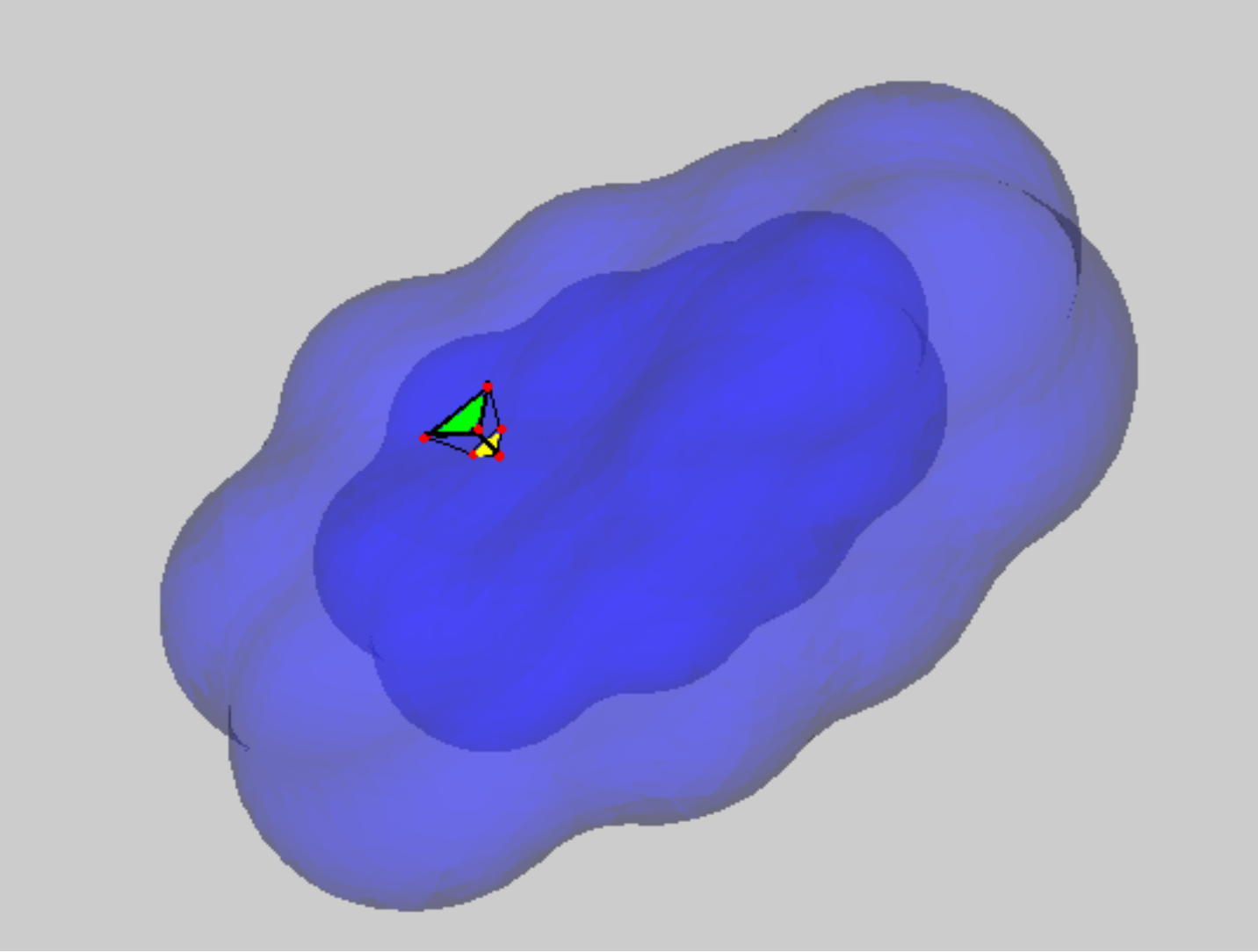

Fig. 10. Transformation of SES into SAS.

Let there be a surface obtained after *primary* rolling only. Then SAS (Fig. 10) is formed from this surface by transferring all surface points along the outer normal at a distance equal to the primary rolling radius $R_{pr}$. In this case, the segments of atomic spheres with the radius $R_{atom}$ are transformed into the segments of SAS spheres with the radius $R_{atom} + R_{pr}$, primary rolling tori degenerate into sphere overlapping lines, and the spherical segmentd of triple points degenerate into the points of overlapping of three or more (in singular cases) spheres.

## 3. Conclusions

The represented algorithm and software developed on the basis if this algorithm allows us to quickly and reliably form a maximally smooth surface over any molecules, including albumins, and then to perform its triangulation. The surface proves to be colored depending on the type of the nearest atoms, thus providing a convenience for visualization, and triangulated, thus allowing us not only to calculate its area and the volume enveloped by this surface, but also to solve the integral equation used in the continual model of a solvent with a rather high precision. The surface obtained via this program may be used both with the purpose of molecular visualization, which is especially topical for large albumin molecules, and for the calculation of the solvation contributions to the intermolecular interaction energy in the presence of a surrounding medium.

## Acknowledgments

The given study is a more detailed presentation and further development of the work [41]. We are profoundly grateful to all the authors for their studies taken as the basis for the given paper.

*УДК 535.71*

# КОНТИНУАЛЬНАЯ МОДЕЛЬ СРЕДЫ I: АЛГОРИТМ ДЛЯ ПОСТРОЕНИЯ ГЛАДКОЙ МОЛЕКУЛЯРНОЙ ПОВЕРХНОСТИ

Купервассер* О.Ю., Ваннер** Н.Э.
*ООО «Транзист Видео», участник Сколково
***Государственное научное учреждение Всероссийский научно-исследовательский институт ветеринарной санитарии, гигиены и экологии Россельхозакадемии, Москва*
*E-mail: olegkup@yahoo.com

В данной работе представлен полный и исчерпывающий алгоритм построения гладкой молекулярной поверхности исключенного объема растворителя - SES, а также поверхности, доступной растворителю - SAS. Эти поверхности играют роль границы между областями молекулы и растворителя. Основой алгоритма является первичная и вторичная обкатки молекул. Оригинальность данной работы состоит в создании *полного и уточненного* алгоритма вторичной обкатки, который позволяет создать оптимально гладкую поверхность SES любой молекулы или системы из молекул, обкатывая любые нерегулярности и близкие к ним ситуации, возникающие при первичной обкатке. Для этого используется адаптивное критическое расстояние, характеризующее максимально допустимую нерегулярность поверхности. Главная задача, которую будет решать полученная поверхность и которая будет рассмотрена в дальнейших статьях, - это расчет энергии сольватации и ее градиентов для континуальных моделей растворителя. Также она может быть использована для демонстрационных целей в молекулярных редакторах.

## 1. Введение.

Цель данной статьи – дать полное и исчерпывающее описание алгоритма, позволяющего построить оптимально гладкую молекулярную поверхность методами первичной и вторичной обкатки, которую мы сможем использовать далее (а) в молекулярных редакторах для демонстрационных целей  или (б) для получения энергии сольватации молекулы  (разница между значениями свободной энергии молекулы в растворе и вакууме) и аналитических градиентов этой энергии. Мы реализуем эту гладкость с помощью первичной обкатки молекулы сферами (внешнюю поверхность мелекулы сфреами с радиусом равным размеру молекулы растворителя) и вторичной обкатки (внутренню поверхность сферой с переменным радиусом, устраня все оставшиеся нерегулярности).

Нужно подчеркнуть особую важность получения гладкой поверхности для этих целей.

Действительно, в случае молекулярного редактора гладкость поверхности необходима для её триангуляции и дальнейшего наглядного изображения, при котором нас не будут отвлекать не имеющие физического смысла нерегулярности.

В случае определения электростатической составляющий энергии сольватации и ее аналитических производных гладкость поверхности является необходимым условием устойчивости алгоритма – фиктивные нерегулярности поверхности приводят к накоплению на них фиктивного заряда, что приводит к неустойчивости алгоритма и большим численным ошибкам [35-37].

Поэтому получение сглаженной поверхности – не только интересная математическая задача, но и практически важная проблема. На основе изложенного в данной работе алгоритма первичной и/или последующей вторичной обкатки были создана программы PQMS[44], MSMS[34], программа Тотрова и Абагяна [20], SIMS[21], TAGSS [38-40] и её усовершенствованная версия, вошедшая как часть в программу DISOLV [39-41]. На основе этих программ и алгоритмов были созданы успешно работающий молекулярный редактор [40] и выдан патент на программу для расчета энергии сольватации и её производных [42].

В чем заключается новизна данной работы по отношению к другим работам [16-19, 20, 21, 34, 38, 44], анализирующим первичную и/или вторичную обкатку молекулярной поверхности? Это, во-первых, полнота – рассмотрены все возможные случаи нерегулярности и показанна возможность их сглаживания. Во-вторых мы стремимся провести сглаживание оптимально. Это означает, что мы стремимся получить не только гладкость, но и в максимальной возможной мере избежать появления участков хоть и уже гладких, но, тем не менее, слишком близких к нерегулярностям (т.е. избегать появленя участков с очень малым радиусом кривизны и очень узких «каналов»). Насколько мы знаем, полностью и исчерпывающее эта задача решена лишь в данной работе.

Однако следует отметить, что без дополнительных ограничений задача построения гладкой поверхности может быть решена тривиально - просто описанием сферы вокруг молекулы. Каковы же эти ограничения?

Во-первых, радиусы атомов это хорошо определенные величины, которые не могут произвольно меняться. В случае использования данного алгоритма для целей изображения молекулы в молекулярном редакторе, радиусы их атомов определяются размерами электронных оболочек. (Это, так называемые, радиусы Ван-дер-Ваальса [1] , [2]. Для радиусов Ван-дер-Ваальса в литературе имеется различные наборы величин [1-3]). В случае использовании их для вычисления энергии сольватации, начальные («огрубленные») значения радиусов задаются как и в первом случае, но затем уточняются таким образом, чтобы вычисленные энергии сольватации молекул соответствовали и их экспериментальным значениям.

Во-вторых, если молекула находится в растворе, то радиус молекул растворителя тоже хорошо определенная величина, которая не может произвольно меняться. Этот радиус безусловно влияет на поверхность раздела между раствором и самой молекулой. Подбираются этото радиус из тех же описаных выше соображений, что и радиус атомов молекул.

Заметим, что указанные выше радиусы достаточно сильно ограничивают возможность варьирования поверхности молекулы. Но, тем не менее, они оставляют достаточно возможностей для того, чтобы сделать её гладкой.

Причем мы хотим сделать её не просто гладкой, но и <u>оптимально</u> гладкой. Под оптимальностью в данном случае понимается, что мы (a) ищем не только гладкую поверхность, но и стремимся уменьшить ее кривизну, не теряя детали поверхности, а также (b) стремимся увеличить декартово расстояние между ***несоседними участками поверхности***. Это участки, близкие пространственно, но удаленные при измерении расстояния между ними вдоль поверхности.  Для оценки степени гладкости задается ***максимальное критическое расстояние***, на которое алгоритмом разрешается приблизиться несоседним участки поверхности. Если это расстояние сближения меньше критического, возникает ситуация близкая к нерегулярности, которая требует вторичной обкатки.

Существует два типа поверхности, окружающей молекулу [4]:

SAS (Solvent Accessible Surface) - поверхность доступная растворителю образуется центрами молекул растворителя, касающихся молекулы субстрата. Число молекул растворителя, касающихся поверхности молекулы пропорционально площади SAS.

SES (Solvent Excluded Surface) - поверхность исключённого из растворителя объёма. Объем, занимаемый растворителем, лежит *вне* объема, ограниченного этой поверхностью. Сам субстрат полностью лежит *внутри* этого объема.

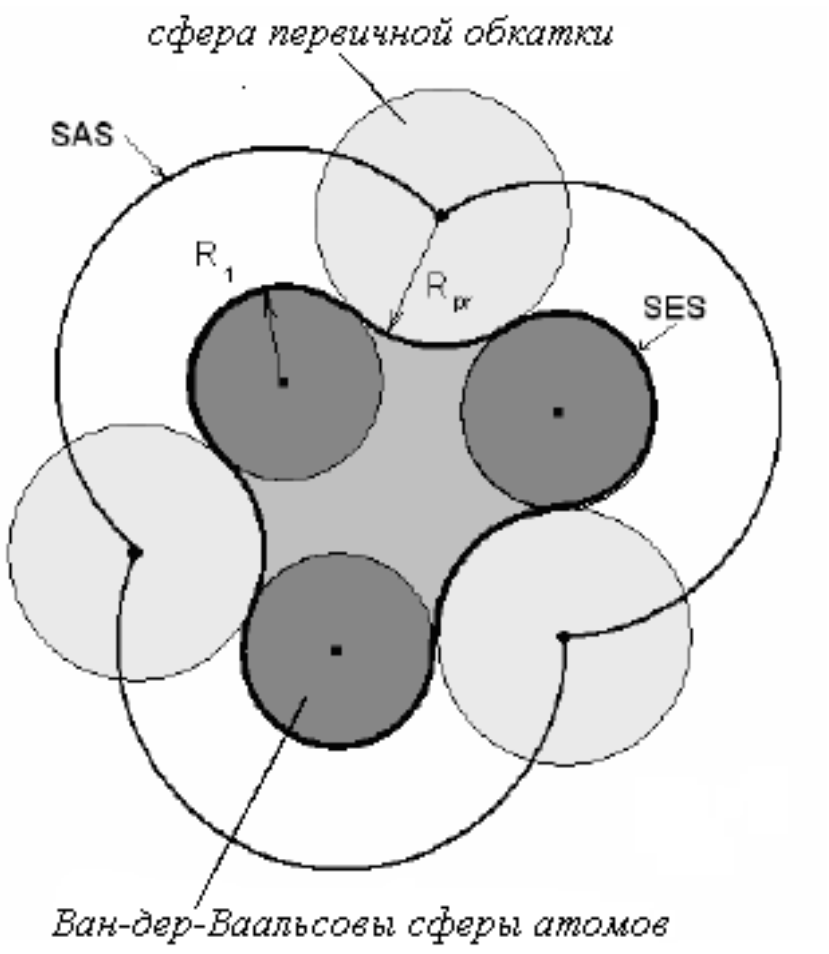

**Рис. 1**. Первичная обкатка поверхности**.** См. [41].

Поверхность SAS можно получить, обкатывая молекулу субстрата молекулой растворителя и отмечая положения ее центра. Обкатка – это движение молекулы растворителя вдоль поверхности субстрата, при котором она последовательно касается всех доступных ей точек субстрата. (Рисунок 1) Для упрощения молекулу растворителя можно заменить сферой обкатки (сфера, описанная вокруг молекулы растворителя). [5-6].

Поверхность SES молекулы можно описывать следующим образом [7]:
(1) сглажено, заменяя ее простыми формами типа [8-15]

(a) сфера

(b) эллипсоид

(c) цилиндр

(2) детально, давая все изгибы поверхности молекулы

(a) покрытием из Ван-дер-ваальсовых сфер вокруг атомов

(b) покрытием из сфер вокруг химических групп атомов

(c) как в предыдущих двух методах, но заполняя остающееся пустое пространство внутри SES фиктивными сферами (это было реализовано в программе GEPOL) [16-19]

(d) Поверхностью уровня для электронной плотности молекулы [22], определяемой через квантовую механику, или иные типы функций, используемые для построения поверхности уровня [23-27]. Данный метод сталкивается с серьёзными трудностями: поверхность уровня может быть также не гладкой; она задаётся не явно, что затрудняет ее триангуляцию; труден подбор функций, дающий поверхность близкую к реальной и определяемой радиусами Ван-де-Ваальса атомов.

(e) соединяя сферы, описанные в (a) или (b) участкам вогнутых и выпуклых поверхностей [20-21].

Наиболее гладкая и реалистичная поверхность получается методом (2)(e), который и рассматривается в этой статье. Получить SES этим методом можно так же как и SAS, обкатывая внешнюю сторону молекулы сферой и беря (а) положения точек касания с атомами (б) отрезки геодезических линий сферы обкатки, проходящими через две точки касания сферы обкатки и атомов (в) сегменты нижней части обкатывающей сферы, ограниченные отрезками геодезических линий сферы обкатки, проходящими через точки касания сферы обкатки и атомов (***первичная обкатка***, Рис. 1). Такое определение поверхности SES было впервые предложено в работе [28]. Предложенный метод получил развитие в работах [20], [29-33]. Однако полученная таким образом поверхность SES может оказаться негладкой [34]. Для дальнейшего сглаживания можно обкатать внутреннюю поверхность молекулы вторично (***вторичная обкатка***, Рис. 2 и 3), что предложено впервые в [21].

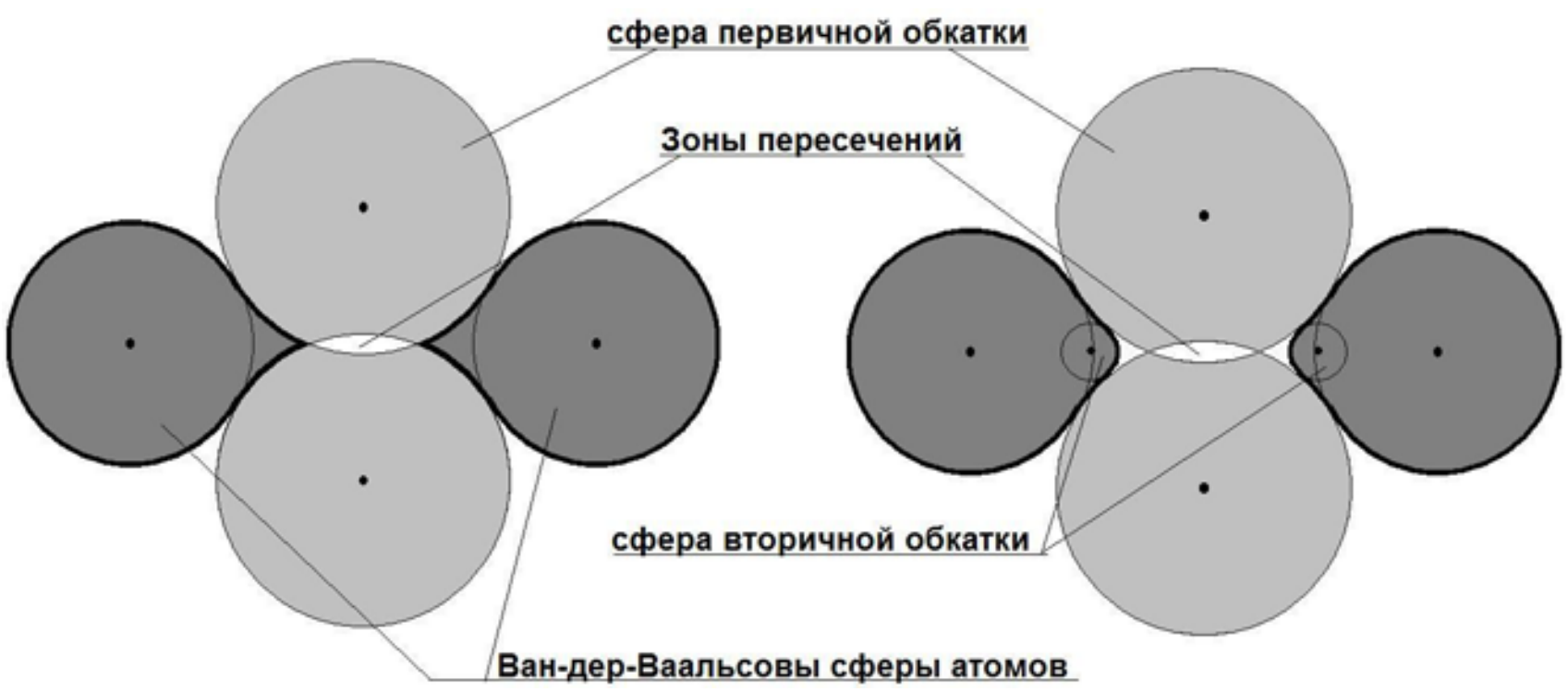

**Рис. 2** Вторичная обкатка поверхности. См.[41].

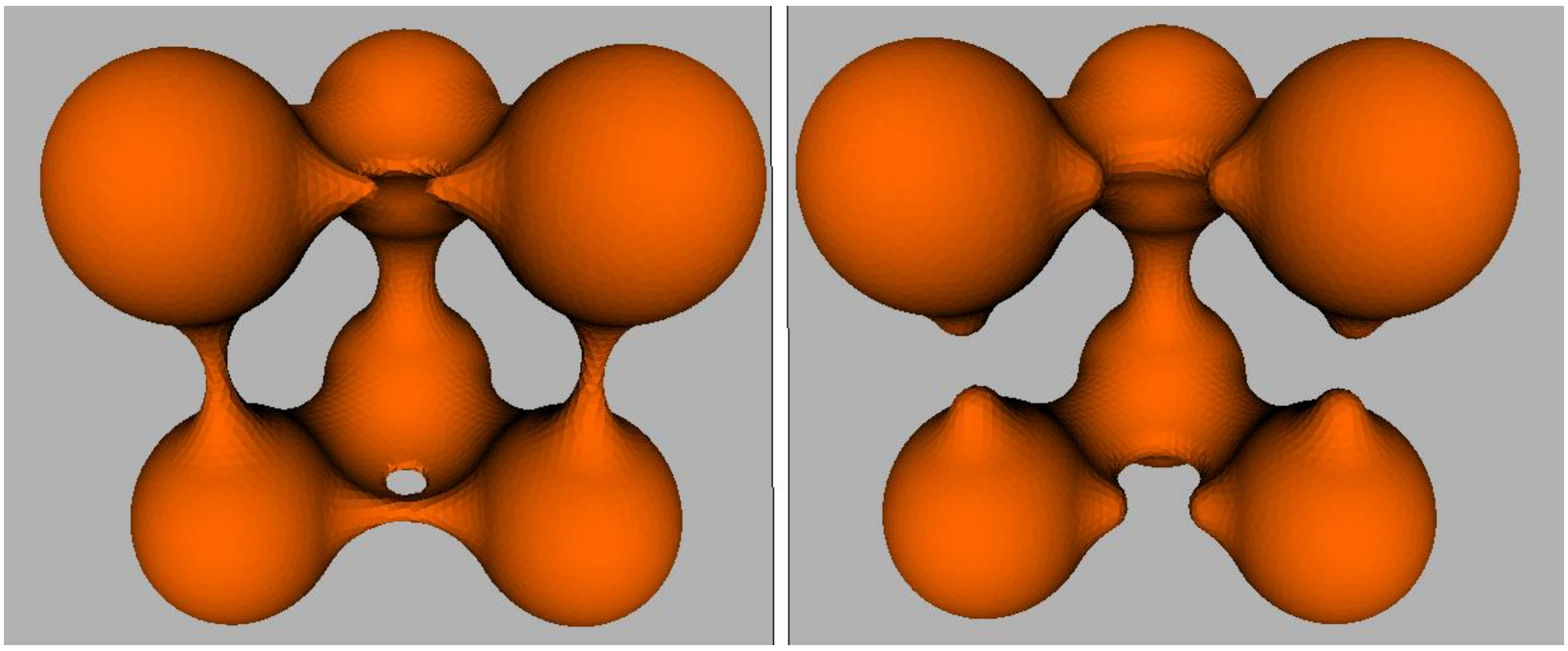

**Рис. 3** Применение метода вторичной обкатки для геометрической конфигурации многих атомов.

Может возникнуть вопрос – почему мы уверены, что вторичная обкатка успешно сгладит все нерегулярности и не потребуется третичная и т.д. обкатки? Аргумент здесь следующий: мы не имели возможность менять радиусы атомов и сферы первичной обкатки. Это заранее заданные фиксированные величины. Радиус же вторичной обкатки – произвольно выбираемая величина, причем даже для обкатки разных участков одной и той же поверхности он может быть выбран разным. Интуитивно ясно, что выбрав радиус вторичной обкатки сколь угодно малым, мы способны «сгладить» все , что угодно. Действительно, в пределе бесконечно малого радиуса, нет проблем для гладкой обкатки любой нерегулярности. Однако мы стремимся не к просто получению гладкой, но оптимально гладкой поверхности. Т.е. она не должна включать элементов хоть и гладких, но близких к нерегулярным. Поэтому следует добиваться вторичной обкатки с радиусом максимально возможным, исходя из геометрических ограничений поверхности. Это делает алгоритм вторичной обкатки достаточно нетривиальным.

Поверхность обкатки состоит из двух типов поверхностных сегментов – сферических и тороидальных. Они подразделяются на фрагменты пяти подтипов. Это сферические элементы ван-дер-ваальсовых сфер атомов, вогнутые сферические элементы первичной обкатки, тороидальные фрагменты первичной и вторичной обкаток обкатки и выпуклые сферические элементы вторичной обкатки. (Рисунок 4)

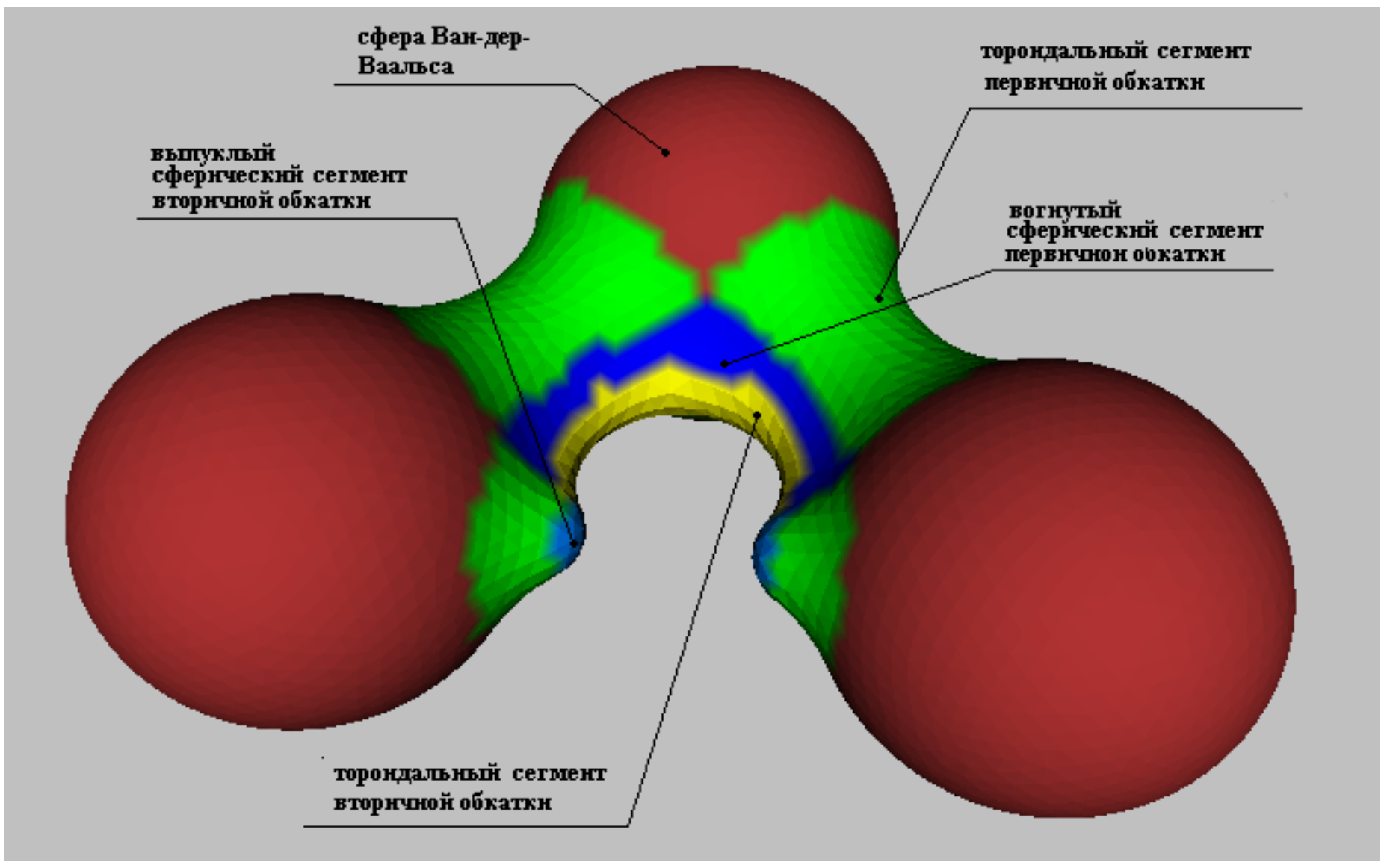

**Рис. 4.** Поверхность молекулы состоит из фрагментов пяти типов. Это сферические элементы ван-дер-ваальсовых сфер атомов, вогнутые сферические элементы первичной обкатки, тороидальные фрагменты первичной и вторичной обкаток обкатки и выпуклые сферические элементы вторичной обкатки.

Радиусы первичной и вторичной обкатки и критическое расстояние имеют ясный физический смысл. Радиус первичной обкатки равен радиусу сферы, описанной вокруг молекулы растворителя. Радиус вторичной обкатки и определенное выше критическое расстояние определяют минимальную кривизну границы молекулы. Нижние пределы для величины радиуса вторичной обкатки и максимального критического расстояния (определенного выше) связаны с «расплыванием» электронных облаков, которые не могут давать слишком острые углы, узкие перешейки и каналы, могущие возникать при первичной обкатке, вследствие соотношения неопределенности Гейзенберга между координатой и импульсом электрона.

В работе [21] (где впервые предложена вторичная обкатка) рассматривается обкатка *не всех типов* нерегулярностей, которые могут возникать при первичной обкатке (b) не рассматривается ситуация «сцепленных» между собой нерегулярностей первичной обкатки (c) отсутствует рассмотрение ситуаций близких к нерегулярным (d) отсутствует решение проблемы нерегулярностей уже вторичной обкатки.

Эти недостатки были преодолены в алгоритме программы TAGSS (Triangulate Area Grid of Smooth Surface) [38-40]. А именно: (a) новый тип нерегулярности, рассмотренный там – пересечение трех первичных сфер (тройная вторичная точка). (b) рассматривались *большие (конфликтные) группы* нерегулярностей, которые обкатывались совместно (c) вторичную обкатку стоит делать и для регулярных участков поверхности, но близкими по свойствам к нерегулярным. Это участки поверхности с малыми декартовыми пространственными расстояниями (меньшие некого *заранее заданного параметра* - критического расстояния) между удаленными, несоседними участками поверхности (т.е. такие участки поверхности, что декартово пространственное расстояние между ними намного больше расстояния вдоль поверхности), узкие каналы или перешейками (d) нерегулярности могут возникнуть уже и для вторичной обкатки. Но убрать их можно просто непрерывно уменьшая радиус сферы вторичной обкатки до тех пор, пока эти нерегулярности не исчезнут.

Однако описанные методы также имели проблемы: (i) если диаметр сферы вторичной обкатки становился меньше критического расстояния, то приходилось принимать определенное выше критическое расстояние за ноль и обкатывать только истинные нерегулярности, меняя радиус вторичной обкатки до необходимой величины. Последнее же приводило к появлению участков со слишком высокой кривизной. (ii) формируемые слишком *большие (конфликтные) группы* нерегулярностей приводили к слишком долгой работе алгоритма и ухудшали гладкость поверхности, повышенная вероятность ситуации, требующей уменьшения радиуса вторичной обкатки (iii) не были полностью рассмотрены все случаи нескольких полостей (а именно, поверхности, соответствующие атомным группам, лежащим внутри крупных полостей, или случай наличия нескольких молекул).

Дальнейшие шаги по преодолению этих проблем были сделаны в [41-43]. Было, соответственно: (i) введено *адаптивное переменное* критическое расстояние, которое могло постепенно уменьшаться в зависимости от типа поверхности (ii) сформулирован корректный и оптимальный механизм формирования *маленьких(конфликтных) групп нерегулярностей*, что позволило повысить гладкость поверхности и быстроту вторичной обкатки (iii) Рассмотрен случай нескольких молекул или случай включения атомных групп в молекулярные полости.

Та часть алгоритма статьи, которая относится ко вторичной обкатке, использует достижения, о которых говорилось выше [41-43] и более подробно описывает, а также развивает их. Такая вторичная обкатка оптимальным образом сглаживает все возможные нерегулярности или близкие к ним ситуации, которые возникают при первичной обкатке. А также разбивает поверхностные атомы и сегменты поверхности на группы, соответствующие разным молекулярным поверхностям, относящимся к разным молекулам или включениям в молекулярные полости.

На основе методов первичной и/или вторичной обкатки  были построены программы , PQMS [44], MSMS [34], программа  Тотрова  и Абагяна [20], SIMS [21], TAGSS [38-40] и её усовершенствованная версия [39-41], вошедшая как часть в программу DISOLV [41-43].  Существуют и более сложные методы получения гладких молекулярных поверхностей, чем механизм вторичной обкатки [22-27, 45-46].

## 2. Этапы построения поверхностей молекул.

### 2.1 Приём входных данных, заполнение внутренних структур. Входные и выходные данные.

Пусть у нас имеется система, состоящая из нескольких молекул в растворителе. Нам известны координаты и типы атомов, входящих в систему и их радиусы Ван-дер-ваальса. Первая стадия алгоритма – определение параметров обкатки (первичной и вторичной). Определение параметров обкатки позволяет получить формальное описание поверхности молекулы в виде наборов координат положения и ориентации сферических и тороидальных фрагментов, а также геометрической связанности этих фрагментов друг с другом. Формируются односвязные поверхности молекул, входящих в систему.  Далее находятся параметры SES и SAS. Входными данными для алгоритма являются массивы координат и Ван-дер-ваальсовых радиусов  атомов. Важными параметрами являются ***радиус первичной сферы обкатки*** и максимальный ***радиус вторичной сферы обкатки***. Задается и ***максимальное критическое расстояние***.Это декартово расстояние, на которое алгоритмом разрешается приблизиться ***несоседним участки поверхности***. Несоседние участки поверхности – это участки удаленые друг от друга при измерении расстояния между ними вдоль поерхности. Если их декартово расстояние сближения меньше критического, возникает ситуация близкая к нерегулярности, которая требует вторичной обкатки.

Радиус первичной обкатки определяется радиусом сферы, описанной вокруг молекулы растворителя. Максимальный радиус вторичной обкатки задается пользователем. Он должен быть меньше половины радиуса первичной обкатки и меньше половины каждого из Ван-дер-ваальсовых радиусов атомов. Максимальное критическое расстояние должно быть меньше максимального диаметра вторичной обкатки. Эти максимальные величины адаптивные и могут уменьшаться самим алгоритмом.

Выходными данными являются массивы параметров, описывающих поверхностные тороидальные и сферические сегменты, поверхностные атомы. Определение параметров обкатки позволяет получить формальное описание гладкой поверхности молекулы в виде наборов координат положения и ориентации сферических и тороидальных фрагментов, а также геометрической связанности этих фрагментов друг с другом.

### 2.2 Определение собственного базиса молекулы и переход в него.

Для простоты рассмотрения мы переходим в собственный базис молекулы. Центр этого базиса находится в центре инерции молекулы. Оси совпадают с главными осями инерции молекулы. Всем атомом при этом вычислении придается одинаковый «вес».

**Радиус-вектор центра инерции**

$$\boldsymbol{r}_c = \frac{1}{N}\sum_{i=1}^{N}(\boldsymbol{r}_{old})_i \quad ,$$

( 1 )

где $N$ – число атомов в молекуле, $(\boldsymbol{r}_{old})_i$ - вектор $i$-того атома.

Тензор, определяющий моменты инерции вычисляется по следующей формуле:

$$F = \frac{1}{N}\sum_{i-1}^{N}\begin{Vmatrix} (r_{old})_{i,x}^2 & (r_{old})_{i,x}(r_{old})_{i,y} & (r_{old})_{i,x}(r_{old})_{i,z} \\ (r_{old})_{i,y}(r_{old})_{i,x} & (r_{old})_{i,y}^2 & (r_{old})_{i,y}(r_{old})_{i,z} \\ (r_{old})_{i,z}(r_{old})_{i,x} & (r_{old})_{i,z}(r_{old})_{i,y} & (r_{old})_{i,z}^2 \end{Vmatrix}. \qquad (2)$$

Диагонализация матрицы $F$ дает нам диагональную матрицу $S$:

$$S = G^T F G.$$

( 3 )

Переход в собственный базис дается формулой:

$$\boldsymbol{r}_i = G((\boldsymbol{r}_{old})_i - \boldsymbol{p}_c)$$

( 4 )

Обратное преобразование дается формулой:

$$\boldsymbol{r}_{old} = G^T \boldsymbol{r} + \boldsymbol{p}_c \quad .$$

( 5 )

## 2.3 Разбиение пространства вокруг и внутри молекулы на перекрывающиеся кубические области.

Разбиение пространства вокруг и внутри молекулы на кубические области представляет собой процесс запоминания индексов атомов молекулы и поверхностных элементов в наборе массивов, каждый из которых соответствует кубической области заданного размера и определённого положения.

Разбиение на кубические области производиться с целью упрощения в дальнейшем процедур поиска индекса атома или элемента поверхности ближайшего к точке с заданными координатами. Кубы строятся и для того, чтобы увеличить скорость перебора поверхностных элементов и атомов при построении поверхности. Слишком удаленные элементы или атомы рассматривать обычно не имеет смысла. Перебор осуществляется только в пределах построенных кубов, что сужает пространство поиска.

Размер кубической области есть постоянный внутренний параметр программы, которые выбирается исходя из известного диапазона возможных значений Ван-дер-ваальсовых радиусов атомов. Расположения и ориентация кубических областей привязаны к осям собственной системы координат. Следует отметить, что кубические области перекрываются между собой. Это позволяет для любой точки или сегмента найти кубическую область таким образом, чтобы эта точка или сегмент лежали вблизи центра этой области.

## 2.4 Первичная и вторичная обкатка

Определение параметров обкатки позволяет получить формальное описание гладкой поверхности молекулы в виде наборов координат положения и ориентации сферических и тороидальных фрагментов, а также геометрической связанности этих фрагментов друг с другом.

### 2.4.1 Первичная обкатка

Первичная обкатка получается катанием шара - зонда первичной обкатки, имитирующего молекулы растворителя, по группе атомов, формирующих молекулу [28]. Она служит для образования молекулярных поверхностей двух типов: SAS и SES. Поверхность SAS описывается положениями центра шар – зонда, имитирующего молекулы растворителя (алгоритм её получения из SES мы рассмотрим в пункте **2.7**), а SES – состоит из нижних фрагментов шара - зонда первичной обкатки [28] в результате обкатки вокруг молекулы. Это (а) положения точек касания с атомами (б) отрезки геодезических линий сферы обкатки, проходящими через две точки касания сферы обкатки и атомов (в) сегменты нижней части обкатывающей сферы, ограниченные отрезками геодезических линий сферы обкатки, проходящими через точки касания сферы обкатки и атомов (Рис. 1).

Поверхность SES первичной обкатки состоит из сферических и тороидальный фрагментов трех типов (Рис.4)

1) В том случае, когда шар первичной обкатки соприкасается только с одним атомом – это выпуклые сферические ***фрагменты Ван-дер-ваальсовых сфер атомов***

2) В том случае, когда шар первичной обкатки соприкасается только с двумя атомами – вогнутые тороидальные фрагменты, образующиеся вращением шара первичной обкатки вокруг этих атомов. Будем называть их ***полные первичные торы*** -если сфера обкатки межет совершить полный круг, не натыкаясь на другие атомы; в ином случае - ***первичными осями***.
3) В том случае, когда шар первичной обкатки соприкасается с тремя атомами (или даже большим числом атомов в вырожденных случаях, например, для ароматических колец) - вогнутые сферические фрагменты, образуемые поверхностью шара первичной обкатки, неподвижно зафиксированного этими атомами (точнее - сегменты нижней части обкатывающей сферы, ограниченные отрезками геодезических линий сферы обкатки, проходящими через точки касания сферы обкатки и атомов). Для наиболее типичного невырожденного случая трех опорных атомов – это вогнутый сферический треугольник. Будем называть их для краткости ***первичными тройными точками***.

Алгоритм, описывающих первичную обкатку следующий:

*1 Шаг:* Находим ***большие первичные конфликтные группы***, которые формируются следующим образом:

Берем любой атом и находим все соседние атомы, лежащие от него на расстоянии менее диаметра первичной обкатки, затем всех их подобных соседей и т.д. Набор всех таких атомов образует первичную большую конфликтную группу. Затем берем любой атом, не вошедший в эту группу, и строим следующую группу. Так до тех пор, пока не исчерпаем все атомы. Первичные большие конфликтные группы образованы несвязанными между собой наборами молекулам. Все большие первичные конфликтные группы обкатываются независимо друг от друга. Делаем это следующим образом:

*2 Шаг:* Находим большие первичные конфликтные группы, состоящие из одного изолированного атома - они допускающие полную обкатку вокруг них без всяких препятствий.

*3 Шаг:* Находим все ***полные первичные торы*** обкатки для текущей большой первичной конфликтной группы, состоящей более чем из одного атома. Они формируются следующим образом:

- Находим все пары атомов из большой группы, которые могут быть на полный круг обкатаны шаром первичной обкатки без того, чтобы он наткнулся при этом на другие атомы.
- Если ширина тора в самом узком месте меньше максимального критического расстояния, определеного выше, или тор самопересекается, то делим тор на две независимые половинки. Это связано с тем, что при самопересечении или дальнейшей вторичной обкатке он распадется на два несвязанных куска.
- Для каждого элемента (атом, полный тор или его половинка) запоминаем его соседние элементы (атом, полный тор или его половинка).

*4 Шаг:* Находим все первичные тройные точки большой группы следующим образом:

- Перебираем все тройки атомов, расстояние между которыми меньше диаметра шара первичной обкатки.
- Для каждой тройки атомов существует либо две тройные точки, либо их нет совсем.

- Если тройные точки существуют, проверяем, что образующая их сфера первичной обкатки не пересекает другие атомы. Подобные тройные точки отбраковываем. В случае просто касания других атомов тройная точка является вырожденной.
- Таким образом, перебираем все возможные тройки атомов.
- Для каждого элемента (атом, первичная тройная точка, полный первичный тор или его половинка) запоминаем его соседние элементы (атом, первичная тройная точка, полный первичный тор или его половинка).

*5 Шаг:* Отбрасываем из рассмотрения в дальнейшем все атомы, не вошедшие в предыдущие группы, поскольку они находятся внутри объема молекулы и не влияют на обкатку. Оставшиеся атомы будем называть поверхностными.

*6 Шаг:* Находим ***малые первичные конфликтные группы***, входящие в больше группы, которые формируются следующим образом:

- Берем любую найденную выше первичную тройную точку.
- Если тройная точка не вырождена, то все ее три опорных атома соединены торами, образовнные полным поворотом сферы обкатки вокруг них. Однако другие атомы могут препятствовать такой полной обкатке. При этом на каждом торе может быть несколько несвязанных сегментов, при обкатке которых первичный шар-зонд упирается в два разных атома. Такие независимые сегменты мы назвали ранее *первичными осями*. Они ограничены двумя крайними положениями сферы первичной обкатки, при которых она упирается в другие атомы.
- Выбираем любой атом, на который опирается исходная тройная точка.
- Находим все тройные точки и все первичные оси, опирающиеся на этот же атом, которые образуются при полной круговой обкатке

вокруг этого атома первичного шара, находящимся первоначально в исходной тройной точке, до тех пор пока при обкатке он не встречает препятствий. Включаем их в малую группу.

- Следует отметить, что тор, пересекаясь с атомом, вырезает из него круглые сегменты. После того, как все торы сделают это «обрезание», оставшаяся необрезанная площадь представляет собой один или несколько несвязанных многоугольных сферических сегментов. Каждый такой атомный сегмент будет тоже рассматривать как независимый элемент. Мы будем вносить такой атомный сегмент в ту малую конфликтную группу, которая определяется опирающимися на этот атомный сегмент тройными точками и осями.
- Если на этот атомный сегмент опирается половинка полного тора – включаем его в малую группу.
- Продолжаем брать опорные атомы тройных точек малой группы до тех пор, пока не исчерпаем их все. Формирование первой малой конфликтной группы закончено.
- Далее берем тройную точку, не вошедшую в уже сформированную малую группу, и формируем новую малую группу. Так до исчерпания всех свободных тройных точек.
- Оставшиеся свободные атомы (вместе с опирающимися на них половинками тора, если таковые есть) также образуют независимые малые группы. Формирование малых конфликтных групп закончено.
- Далее перебираем все пары атомов, соединенные полным тором. Если оба сегмента атомов, на который опирается полный тор, принадлежат одной и той же малой группе – включаем тор в эту малую группу. Если разным малым группам – объединяем эти две малые группы в одну и включаем в нее этот полный тор.
- Для каждого элемента (атом, тройная точка, полный тор или его половинка, сферические и тороидальные сегменты) запоминаем его соседние элементы.

*7 Шаг:* Фильтрация «паразитных» поверхностей.

Полученные малые первичные конфликтные группы включают одну или несколько независимых замкнутых поверхностей, которые образуются при первичной обкатке. Это:

1) Внешняя поверхность молекулы.
2) Внутренняя полость, попав в которую, шар первичной обкатки не может выбраться наружу. Внутренние полости обычно имеют нормаль, направленную внутрь полости. Следовательно, написав формулу для нахождения объема полости $V = \frac{1}{3}\oint_S (\boldsymbol{r} \cdot \boldsymbol{n})dS$ можно видеть, что объем в этом случае получается отрицательным. Это позволяет отфильтровать такие полости
3) Внешние поверхности группы атомов, «запертых» во внутреннюю полость. Эту поверхность можно отфильтровать по следующим признакам – ее внутренние атомы являются также и внутренними атомами для внешней поверхности молекулы. При этом средний радиус внешней поверхности больше.
4) Случай, когда группы атомов могут распадаться на несколько несвязанных  молекул (его обработка расссмотрена в пункте **2.5**). Это происходит, когда первичные оси разрываются при самопересечении или вторичной обкаткой. Для полной обработки таких ситуаций необходимо вначале провести вторичную обкатку.

Процедура первичной обкатки иногда может приводить к нежелательным самопересечениям поверхности и изломам SES. Эти нерегулярности можно разделить на два типа, причем нерегулярность может быть одновременно образована несколькими элементами разного или одинакового типа:

1) Самопересечение тора первичной обкатки.

2) Пересечение вогнутых сферических фрагментов первичной обкатки – излом.

При этом могут возникать ситуации хоть и гладкие, но близкие к этим нерегулярным элементам поверхности. Это происходит, когда

1) ширина тора в самом узком месте меньше некого заранее эмпирически выбранного критического расстояния

2) расстояние между вогнутыми сферическими сегментами меньше этого же критического расстояния.

Для преодоления этих проблем применяется метод вторичной обкатки [21,38-43].

### 2.4.2 Вторичная обкатка.

Мы выбираем шар-зонд вторичной обкатки с диаметром, меньшим половины радиуса первичной обкатки и меньшим половины радиусов всех атомов. Этим шаром мы обкатываем *внутреннюю* поверхность молекулы, полученную в результате первичной обкатки, только в тех местах, где возникают нерегулярности или близкие к нерегулярности элементы поверхности.

Но обкатывая эти места, мы неизбежно должны обкатать и все связанные с ними другие фрагменты первичной обкатки. Так, например, обкатывая шаром вторичной обкатки пересечение вогнутых сферических сегментов первичной обкатки, мы неизбежно должны обкатать и торы первичной обкатки, которые с ними соприкасаются. Другие возможные ситуации будут описаны ниже. Так образуются наборы фрагментов первичной обкатки, которые нуждаются во вторичной обкатке и объединяются в *конфликтные группы вторичной обкатки.*

При обкатке таких конфликтных групп могут возникать новые нерегулярности (или близкие к ним ситуации), связанные с пересечением или сближением на расстояние меньшего критического фрагментов шаров вторичной обкатки. Однако, третичная обкатка уже не нужна! Мы можем ликвидировать эти новые нерегулярности (или близкие к ним ситуации) путем плавного уменьшением радиуса вторичной обкатки до той величины, когда эти проблемы исчезают. Также может возникнуть необходимость уменьшения критического расстояния, определяющего ситуации близкие к нерегулярным. Эта необходимость возникает при уменьшении диаметра шара вторичной обкатки (проводимого для ликвидации этих новых нерегулярностей или близких к ним ситуаций) до величины меньшей этого критического расстояния. Диаметр шара вторичной обкатки должен оставаться всегда больше этого критического расстояния.

Поверхность SES вторичной обкатки, сглаживающая нерегулярность, состоит из сферических и тороидальный фрагментов трех типов Рис. 4.

1) В том случае, когда шар вторичной обкатки соприкасается только с двумя вогнутыми сферическими сегментами первичной обкатки – выпуклые тороидальные фрагменты вторичной обкатки, образующиеся вращением шара вторичной обкатки вокруг этих сферических сегментов – вторичные торы. Сегменты вторичных торов называются вторичными осями.
2) В том случае, когда шар первичной обкатки соприкасается с тремя (и более в вырожденных случаях) сферическими сегментами первичной обкатки - выпуклые сферические фрагменты, образуемые поверхностью шара вторичной обкатки, неподвижно зафиксированного этими первичными сферами – вторичные тройные точки.
3) В случае, когда образуется самопересечение тора первичной обкатки или близкая к этому ситуация узкого перешейка тора – возникает пара устойчивых положений шара вторичной обкатки вблизи пары точек самопересечения или узкого перешейка тора. Выпуклые сферические сегменты шара вторичной обкатки, сглаживающие нерегулярность образует третий тип – пары точек устойчивого положения. Смотри Рис. 2.

Мы видим, что первые два типа сглаживания аналогичны типам, возникающим при первичной обкатке, и появляются при вторичной обкатке первичных сфер. Третий тип новый и возникает вследствие обкатки первичных торов, которых не было при первичной обкатке.

На Рис. 2, 3, 4 показана процедура сглаживания изломов и удаления области самопересечения при помощи метода вторичной обкатки. Она происходит следующим образом. Для каждой найденной ранее малой *первичной* конфликтной группы вторичная обкатка производится независимо.

1) Обкатываем полные торы, с шириной перешейка меньше критического или самопересекающиеся. Образуется пара точек устойчивого положения. Вторичная конфликтная группа не возникает.

2) Обкатываем вторичным шаром-зондом пересекающиеся или лежащие на расстоянии меньше критического первичные тройные точки. Если ничто не препятствует обкатке, образуется полный вторичный тор и вторичная конфликтная группа не возникает.

3) В случаях, отличных от двух предыдущих, возникают сцепленные нерегулярности и образуются их вторичные конфликтные группы. Формируем вторичные конфликтные группы первичных тройных точек следующим образом: Берем любую первичную тройную точку и находим все соседние первичные тройные точки, лежащих от нее на расстоянии менее диаметра вторичной обкатки, затем все их подобные соседи и т.д. Комплекс всех таких первичных тройных точек образует *большую вторичную конфликтную группу*. Затем берем любую первичную тройную точку, не вошедшую в эту группу, и строим следующую группу. Так до тех пор, пока не исчерпаем все первичные тройные точки. Внутри таких *больших вторичных конфликтных групп* будем искать *малые вторичные конфликтные группы* связанных нерегулярностей или близких к нерегулярностям ситуаций.

4) Обкатываем вторичным шаром-зондом пересекающиеся или лежащие на расстоянии Δ меньше критического две первичные тройные точки. Начинаем с пересекающихся тройных точек (или с минимального Δ при их отсутствии). (Далее для формирования следующих малых вторичных конфликтных групп берем тройные точки в порядке увеличения Δ.) Запоминаем это текущее расстояние Δ (при отсутствии пересечения) и радиус вторичной обкатки в массив «резервных» критических величин. При обкатке образуется вторичные оси. Движению вторичного шара-зонда вдоль этих осей может что-то препятствовать.

   - Движению вдоль этой вторичной оси может помешать другая первичная тройная точка из большой вторичной конфликтной группы (В принципе, может быть и вырожденный случай более одной такой первичной тройной точки, *одновременно* останавливающих обкатку). При этом образуется уже вторичная тройная точка, опирающаяся на три первичные тройные точки. Также образуются новые, выходящие из нее вторичные оси. Включаем пары первичных тройных точек (служащих опорой для этой вторичной тройной точки и для исходящих из нее вторичных осей) в малую вторичную конфликтную группу для их дальнейшей обкатки.
   - Торы первичной обкатки из большой конфликтной группы, примыкающие к этой паре первичных тройных точек, могут служить препятствием для обкатки вторичного тора. Включаем эти торы и связанные с ними первичные оси в малую вторичную конфликтную группу для их дальнейшей обкатки.

- При обкатке такого тора образуется пара точек устойчивого положения. Мы должны обкатать и все пары первичных тройных точек, примыкающие к такому тору. Включаем эти пары первичных тройных точек в малую вторичную конфликтную группу для их дальнейшей обкатки.
- Обкатываем все (как уже запомненные ранее, так и образующиеся по тому же принципу снова) пары первичных тройных точек и первичных осей (входящие в формируемую малую конфликтную группу) до их полного исчерпания.

5) Может возникнуть ситуация, когда при вторичной обкатке, описанной в пункте 4:

а. Могут попасть под вторичную обкатку нерегулярность или близкая к ней ситуация, *уже* входящая в другую, ранее образованную малую конфликтную группу.

б. Образуется вторичная ось с перешейком, меньше критического (ширина $\Delta$) или самопересекающаяся..

в. Пара сфер устойчивого положения пересечется или сблизится на расстояние ($\Delta$), меньшее критического.

г. Пара вторичных тройных точек пересечется или сблизится на расстояние ($\Delta$), меньшее критического.

Тогда уменьшаем непрерывно диаметр вторичной обкатки до того максимально большого диаметра $D_{new}$, когда описанная выше ситуация с обкаткой ранее обкатанной нерегулярности, либо нерегулярность или близкая к ней ситуация исчезает. Далее возвращаемся в начало пункта 4 и формируем заново конфликтную группу. При этом критическое расстояние и диаметр вторичной обкатки выбираем следующим образом:

b. Пусть этот диаметр $D_{new}$ *больше* текущего критического расстояния. Тогда выбираем новый диаметр обкатки равный ему , а критическое расстояние оставляем прежним. Если минимальное расстояние $\Delta$ ,

приведшее  к проблеме, больше нуля, то запоминаем величину этого расстояния  и найденный радиус вторичной обкатки в массив «резервных» критических расстояний.

c. Пусть этот диаметр $D_{new}$ *меньше* текущего критического расстояния и максимальное «резервное» критическое расстояние *больше* этого нового диаметра вторичной обкатки. Тогда выбираемое заново критическое расстояние равно максимальному «резервному», а радиус вторичной обкатки – также «резервный» соответствующий выбранному критическому расстоянию. Из «резерва» выбрасываются эти величины.
d. Пусть этот диаметр $D_{new}$ *меньше* текущего критического расстояния и максимальное «резервное» критическое расстояние *меньше* этого нового диаметра вторичной обкатки. Тогда выбираемые заново критическое расстояние и диаметр вторичной обкатки берутся равными этому  диаметру $D_{new}$ вторичной обкатки.
e. Если одно из найденных критических расстояний слишком мало (например, меньше допускаемого машинной точностью) увеличиваем радиусы атомов на 0.1 А и повторяем первичную и вторичные обкатку.

6) Продолжаем это процесс до тех пор, пока не будут сформированы все малые вторичные конфликтные группы и определены критические расстояния и диаметры вторичной обкатки для них. Для каждого элемента (атом, тройная точка, полный тор или его половинка, первичные и вторичные сферические и тороидальные сегменты) запоминаем его соседние элементы.

## 2.5 Формирование независимых групп поверхностных атомов и сегментов, каждая из которых соответствует одной и только одной поверхности.

Здесь мы даем алгоритм для случая, когда группы атомов могут распадаться на несколько несвязанных молекул (Это, например, случай докинга лиганда к субстрату ). При этом происходит формирование независимых групп поверхностных атомов и сегментов, каждая из которых соответствует одной и только одной поверхности.

Малые *первичные* конфликтные группы могут распасться на подгруппы, соответствующие разным поверхностям. Это происходит, когда первичные оси разрываются при самопересечении или сферами устойчивого положения вторичной обкатки.

Однако такой разрыв однозначно происходит лишь только для такого типа первичной оси, где две ее смежные первичные тройные точки (ограничивающие ее угловые размеры) лежат на трех или более одинаковых опорных атомах. Другие типы первичных осей могут не разорваться, поскольку оказываются связанными возникшей после вторичной обкатки *сетью* из одной или более вторичных осей. Мы устанавливаем новые связи атомов через эту вновь возникшую сеть.

Делая анализ связанность (т.е. разбивая атомы малых первичных конфликтных групп на другие подгруппы из-за разрыва первичных осей (торов) и иногда восстанавливая связанность этих подгрупп через сеть вторичных торов), находим те первичные группы поверхностных атомов и сегментов, которые будут соответствовать разным поверхностям.

## 2.6 Определение параметров сферических и тороидальных сеггментов

### 2.6.1 Определение параметров выпуклых сферических сегментов поверхностных атомов.

Выпуклые сферические фрагменты являются частями Ван-дер-ваальсовых сфер атомов, которые обращены к растворителю. Для описания этих сферических фрагментов для каждого атома используется набор векторов описывающих направления на соседние контактные атомы и набор значений «запрещённых» углов связанных с каждым из направлений. Контактными считаются атомы, имеющие расстояние между своими сферами меньше диаметра сферы первичной обкатки $2R_{pr}$. А «запрещённый» угол – это угол, под которым из центра атома виден радиус дуговой траектории центра сферы обкатки при касании пары атомов – текущего и второго контактного. Та часть Ван-дер-ваальсовской сферы данного атома, которая лежит вне всех «конфликтных конусов» является искомым выпуклым сферическим фрагментом поверхности молекулы. Следует отметить, что запрещенные углы могут разбивать атомную сферу не только на один, но и на два или даже более несвязанных между собой сферических фрагментов $S_{R_j}$.

Пусть $\boldsymbol{r}_0$ - вектор координат текущего атома, а $\boldsymbol{r}_i$ - вектор координат контактного атома; $R_0$ - радиус текущего атома, а $R_i$ - радиус контактного атома;

Для каждого атома определены массивы единичных векторов $\vec{d}_i$ и массивы запрещённых углов: $\gamma_i$. Вектор направления определяется из условия:

$$\boldsymbol{d}_i = \frac{\boldsymbol{r}_i - \boldsymbol{r}_0}{|\boldsymbol{r}_i - \boldsymbol{r}_0|},$$

( 6 )

а запрещённый угол:

$$cos(\gamma_i) = \frac{b^2 - a^2 - c^2}{2ac}, \quad a = R_0 + R_{pr}, \quad b = R_i + R_{pr}, \quad c = |\boldsymbol{r}_i - \boldsymbol{r}_0|, \qquad (7)$$

где индексом ‘0’ обозначен текущий атом, а индексом ‘*i*’- контактный атом. Алгоритм определения параметров выпуклых сферических элементов состоит в переборе всех пар контактных атомов и вычислении для каждого из атомов набора направляющих векторов и соответствующих запрещённых углов.

### 2.6.2 Определение массива координат центров первичных вогнутых сферических фрагментов (первичная сферическая перетяжка) между тройками атомов.

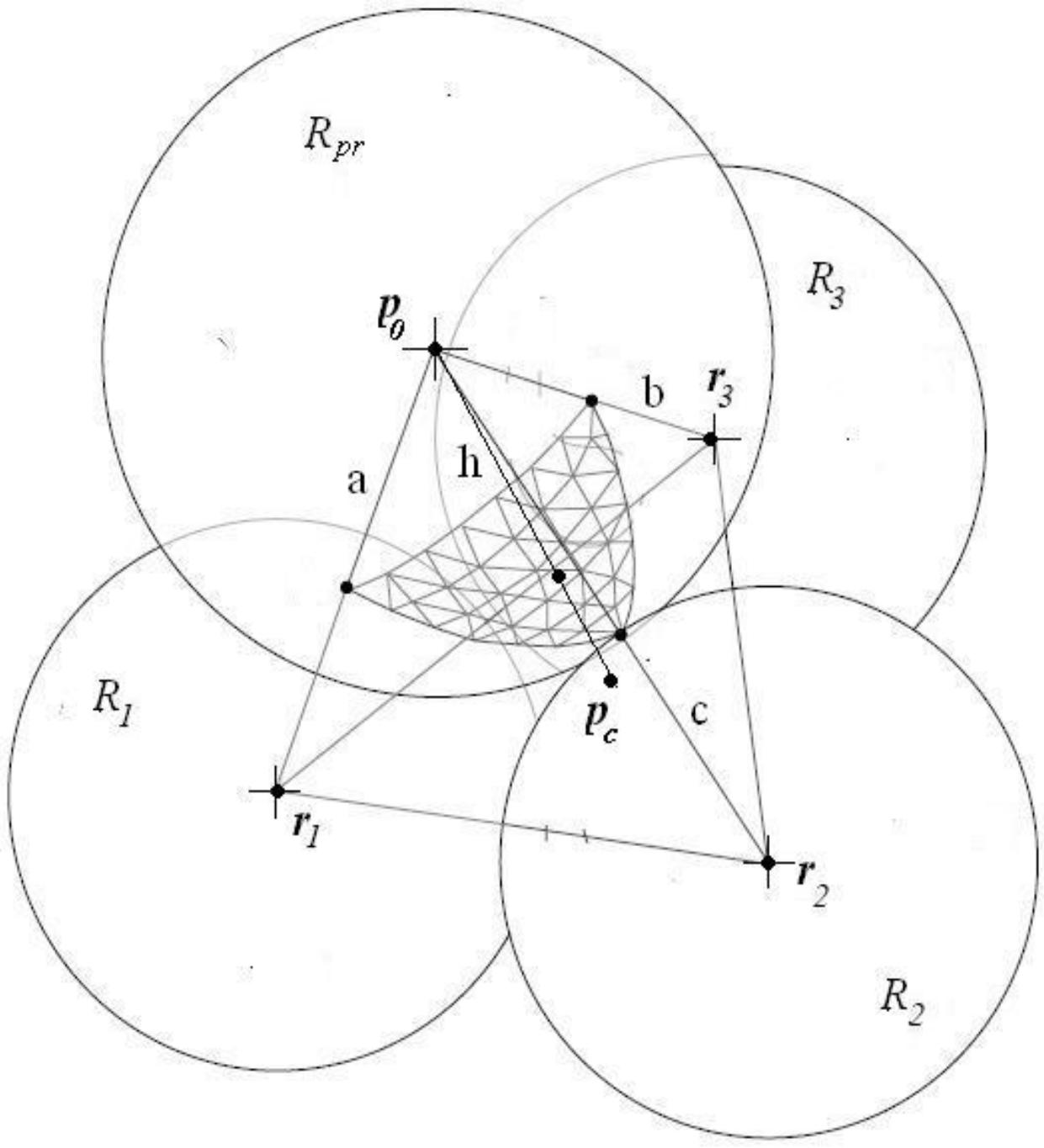

**Рис. 5.** Геометрия задачи об определении центра сферы касающейся трёх атомов.

Вогнутые сферические элементы поверхности SES (Рис. 5) имеют форму сферического треугольника и образуются при контакте сферы обкатки одновременно с тремя атомами.(Кроме редких вырожденных случаев. При этом точки контактов сферы обкатки со сферами атомов определяют вершины сферического треугольника являющегося частью сферы обкатки.

Расчёт координат вектора $\boldsymbol{p}_0$ центра сферы $R_{pr}$ опирающейся на три атомами с координатами $\boldsymbol{r}_1, \boldsymbol{r}_2, \boldsymbol{r}_3$ производится по нижеследующим формулам.

Боковые стороны пирамиды:

$$a = R_1 + R_{pr}; b = R_2 + R_{pr}; c = R_3 + R_{pr}, \qquad (8)$$

где $h$ - высота пирамиды, $\boldsymbol{p}_c$ - радиус-вектор основания высоты пирамиды.

$$\boldsymbol{z} = \frac{[(\boldsymbol{r}_3 - \boldsymbol{r}_1) \times (\boldsymbol{r}_2 - \boldsymbol{r}_1)]}{\|[(\boldsymbol{r}_2 - \boldsymbol{r}_1) \times (\boldsymbol{r}_3 - \boldsymbol{r}_1)]\|}, \qquad (9)$$

$$\boldsymbol{r} = \boldsymbol{p}_c \pm \boldsymbol{z} h, \qquad (10)$$

$$\begin{cases} (\boldsymbol{r} - \boldsymbol{r}_1)^2 = a^2 \\ (\boldsymbol{r} - \boldsymbol{r}_2)^2 = b^2 \\ (\boldsymbol{r} - \boldsymbol{r}_3)^2 = c^2 \end{cases} \Leftrightarrow \qquad (11)$$

$$\begin{cases} (\boldsymbol{p}_c - \boldsymbol{r}_1)^2 = a^2 - h^2 \\ (\boldsymbol{p}_c - \boldsymbol{r}_2)^2 = b^2 - h^2 \\ (\boldsymbol{p}_c - \boldsymbol{r}_3)^2 = c^2 - h^2 \end{cases} \Rightarrow \qquad (12)$$

$$\begin{cases} (\boldsymbol{r}_2 - \boldsymbol{r}_1) \cdot \boldsymbol{p}_c = (a^2 - b^2 - \vec{r}_1^{\,2} + \vec{r}_2^{\,2}) / 2, \\ (\boldsymbol{r}_3 - \boldsymbol{r}_1) \cdot \boldsymbol{p}_c = (a^2 - c^2 - \vec{r}_1^{\,2} + \vec{r}_3^{\,2}) / 2 \end{cases} \qquad (13)$$

$$\boldsymbol{p}_c = \boldsymbol{r}_1 + \alpha[-\boldsymbol{z} \times (\boldsymbol{r}_2 - \boldsymbol{r}_1)] + \beta[\boldsymbol{z} \times (\boldsymbol{r}_3 - \boldsymbol{r}_1)], \qquad (14)$$

$$\boldsymbol{p}_c = \boldsymbol{r}_1 + [\boldsymbol{z} \times (\gamma \boldsymbol{r}_1 - \alpha \boldsymbol{r}_2 + \beta \boldsymbol{r}_3)], \qquad (15)$$

$$\begin{cases} \alpha = \dfrac{(a^2 - c^2 + (\boldsymbol{r}_3 - \boldsymbol{r}_1)^2)}{2\|[(\boldsymbol{r}_2 - \boldsymbol{r}_1) \times (\boldsymbol{r}_3 - \boldsymbol{r}_1)]\|} \\ \beta = \dfrac{(a^2 - b^2 + (\boldsymbol{r}_2 - \boldsymbol{r}_1)^2)}{2\|[(\boldsymbol{r}_2 - \boldsymbol{r}_1) \times (\boldsymbol{r}_3 - \boldsymbol{r}_1)]\|} \\ \gamma = \alpha - \beta = \dfrac{(b^2 - c^2 - (\boldsymbol{r}_2 - \boldsymbol{r}_1)^2 + (\boldsymbol{r}_3 - \boldsymbol{r}_1)^2)}{2\|[(\boldsymbol{r}_2 - \boldsymbol{r}_1) \times (\boldsymbol{r}_3 - \boldsymbol{r}_1)]\|} \end{cases}, \qquad (16)$$

$$[(r_3-r_1)\times(r_2-r_1)]=[(r_2-r_3)\times(r_1-r_3)]=[(r_1-r_2)\times(r_3-r_2)], \qquad (17)$$

$$p_c=\frac{(r_1+r_2+r_3)}{3}+\frac{1}{6}\left[\left(\begin{array}{l}\frac{3(b^2-c^2)+(r_3-r_1)^2-(r_2-r_1)^2}{\|[(r_3-r_1)\times(r_2-r_1)]\|^2}[[(r_3-r_1)\times(r_2-r_1)]\times r_1]+\\+\frac{3(c^2-a^2)+(r_1-r_2)^2-(r_3-r_2)^2}{\|[(r_1-r_2)\times(r_3-r_2)]\|^2}[[(r_1-r_2)\times(r_3-r_2)]\times r_2]+\\+\frac{3(a^2-b^2)+(r_2-r_3)^2-(r_1-r_3)^2}{\|[(r_2-r_3)\times(r_1-r_3)]\|^2}[[(r_2-r_3)\times(r_1-r_3)]\times\vec{r}_3]\end{array}\right)\right], \qquad (18)$$

$$h^2=\frac{a^2+b^2+c^2-(r_1^{\ 2}+r_2^{\ 2}+r_3^{\ 2})+2p_c\cdot(r_1+r_2+r_3)}{3}-p_c^{\ 2}\ . \qquad (19)$$

Для каждой сферы первичной обкатки определены массивы единичных векторов $d_i$ и массивы запрещённых углов: $\gamma_i$. Они связанны с вторичными торами обкатки вогнутых сферических элементов. Вектор направления определяется из условия:

$$d_i=\frac{r_i-p_0}{|r_i-p_0|}, \qquad (20)$$

а запрещённый угол:

$$cos(\gamma_i)=\frac{b'^2-a'^2-c'^2}{2a'c'},\quad a'=R_{pr}+R_{sec},\quad b'=R_{pr}+R_{sec},\quad c'=|r_i-p_0|, \qquad (21)$$

где индексом ‘0’ обозначена текущая первичная сфера, а индексом ‘$i$’- контактная первичная сфера.

Алгоритм определения параметров вогнутых сферических элементов состоит в переборе всех пар вогнутых сферических элементов, для которых проводится вторичная обкатка, и вычислении для каждого из вогнутых сферических элементов набора направляющих векторов и соответствующих запрещённых углов.

### 2.6.3 Определение массивов параметров тороидальных фрагментов между парами атомов.

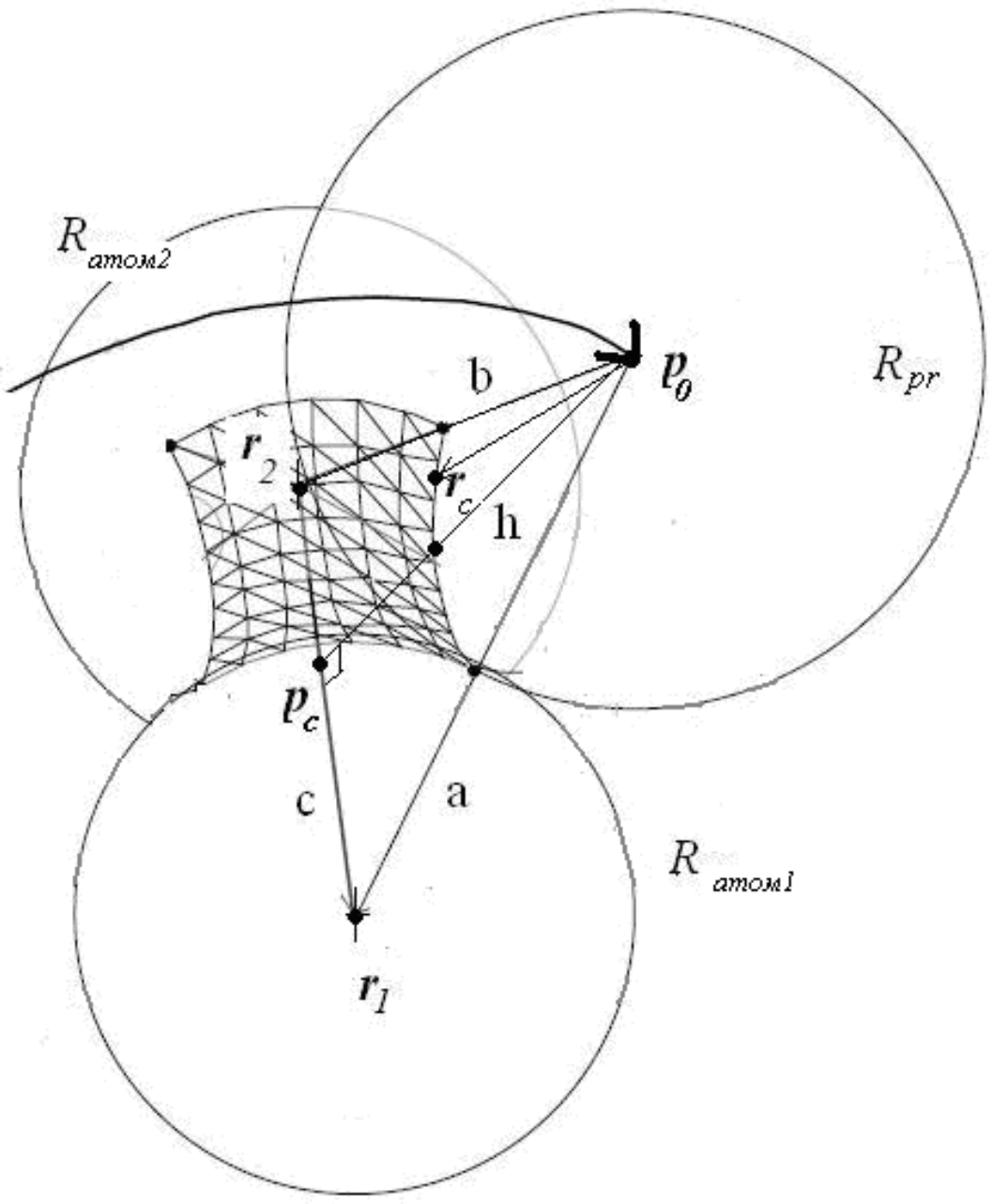

**Рис. 6.** Обкатка двух атомов первичным шаром-зондом. Формирование первичного тора.

Первичные тороидальные фрагменты поверхности (Рис. 6) образуются в процессе обкатки первичной сферой обкатки двух атомов. При этом центр сферы обкатки описывает круговую дугу или полную окружность в зависимости от расположения других атомов. Тороидальный фрагмент заключён между точками касания сферы двух атомов и положениями сферы обкатки, в которых она упирается в какой-либо другой атом, то есть положениями касания сферы обкатки одновременно трёх атомов.

Для описания тороидальных фрагментов используются следующие параметры.

Центр $\boldsymbol{p}_c$ и радиус $h$ окружности, вокруг которой описывает дугу центр сферы обкатки.

Локальный базис $x,y,z$, ось $z$ которого направлена параллельно прямой соединяющей центры опорных атомов, а оси $x$ и $y$ определяют плоскость дуги центра сферы обкатки, при этом ось $x$ направлена на один из атомов, который ограничивает обкатку вокруг данной пары атомов.

Углы $\alpha$ и $\beta$ относительно оси $x$ начала и конца дуговой траектории центра сферы обкатки.

Угол γ, описывает максимальное угловое отклонение от прямой, определяемой осью z, радиус-вектора из текущей точки тора в текущий центр сферы обкатки $\boldsymbol{p}_0$- $\boldsymbol{r}_s$. При этом отклонении она ещё не вырезается из тора одной из сфер устойчивого положения, если была необходимость во вторичной обкатке тора

Для элементов находящихся на тороидальной перетяжке между атомами $\boldsymbol{r}_1$ и $\boldsymbol{r}_2$:

$R_{pr}$-радиус первичной сферы обкатки, $R_{sec}$-радиус вторичной сферы обкатки, $R_1$ – радиус первого атома, $R_2$ – радиус второго атома

Промежуточные расчёты проводятся по нижеследующим формулам:

$$a = R_1 + R_{pr}; b = R_2 + R_{pr}; c = \left|\boldsymbol{r}_2 - \boldsymbol{r}_1\right|, \qquad (22)$$

$c$- расстояние между атомами.

Радиус круговой траектории центра сферы обкатки:

$$h = \frac{1}{2c}\sqrt{4a^2c^2 - \left(a^2 + c^2 - b^2\right)^2} . \qquad (23)$$

Центр круговой траектории:

$$\boldsymbol{p}_c = \tfrac{1}{2}\left(\boldsymbol{r}_1 + \boldsymbol{r}_2\right) + \frac{\left(\boldsymbol{r}_2 - \boldsymbol{r}_1\right)\left(a^2 - b^2\right)}{2c^2} . \qquad (24)$$

Локальный ортонормированный базис

$$\boldsymbol{z} = \frac{\left(\boldsymbol{r}_2 - \boldsymbol{r}_1\right)}{c}, \quad (\boldsymbol{x}\cdot\boldsymbol{z}) = 0, \quad (\boldsymbol{y}\cdot\boldsymbol{z}) = 0, \quad (\boldsymbol{x}\cdot\boldsymbol{y}) = 0 . \qquad (25)$$

Если атом с координатами $\vec{r}_3$ ограничивает обкатку вокруг атомов $\vec{r}_1$ и $\vec{r}_2$, то определение базисных векторов задано выражениями:

$$\boldsymbol{l} = \frac{\boldsymbol{r}_3 - \boldsymbol{p}_c}{\left|\boldsymbol{r}_3 - \boldsymbol{p}_c\right|}, \quad \boldsymbol{y} = [\boldsymbol{z}\times\boldsymbol{l}], \quad \boldsymbol{x} = [\boldsymbol{z}\times\boldsymbol{y}], \qquad (26)$$

Расчёт углов $\alpha$ и $\beta$ производится по формулам:

$$\alpha = arccos\left( \boldsymbol{x} \cdot \frac{\boldsymbol{p}_{0,\alpha} - \boldsymbol{p}_c}{\left| \boldsymbol{p}_{0,\alpha} - \boldsymbol{p}_c \right|} \right), \quad \beta = arccos\left( \boldsymbol{x} \cdot \frac{\boldsymbol{p}_{0,\beta} - \boldsymbol{p}_c}{\left| \boldsymbol{p}_{0,\beta} - \boldsymbol{p}_c \right|} \right), \qquad (27)$$

где $\boldsymbol{p}_{0,\alpha}$ и $\boldsymbol{p}_{0,\beta}$ - координаты центра сферы обкати при касании трёх атомов на начале и конце дуги обкатки вокруг пары атомов. Расчёт угла γ производится по формуле

$$sin(\gamma) = \frac{h}{R_{pr} + R_{sec}} \quad . \qquad (28)$$

### 2.6.4 Определение массива координат центров и радиусов вторичных сферических фрагментов (вторичная сферическая перетяжка) между тройками первичных сферических фрагментов .

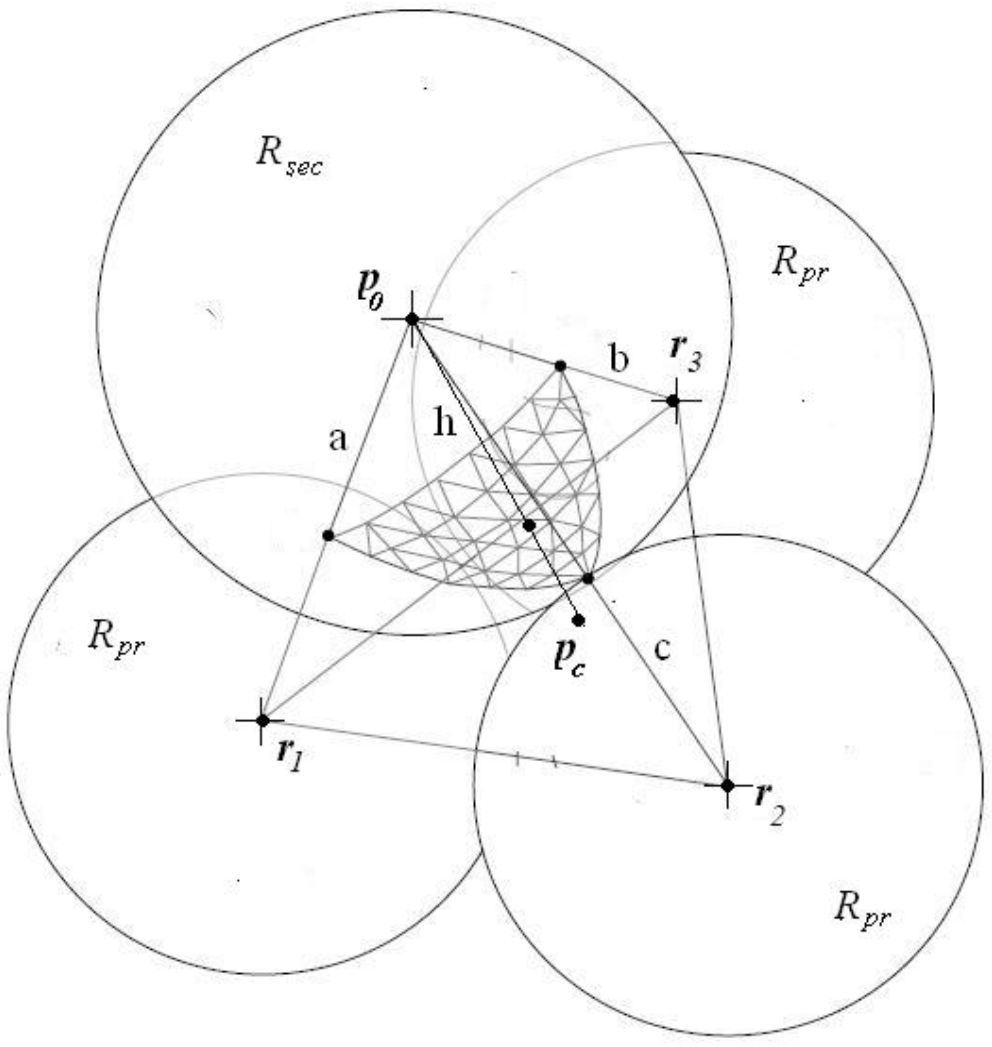

**Рис. 7.** Геометрия задачи об определении центра вторичного шара-зонда, касающегося трёх первичных сфер обкатки.

Расчёт координат вектора $\boldsymbol{p}_0$ центра вторичной сферы опирающейся на три первичные сферы обкатки (Рис. 7) с координатами $\boldsymbol{r}_1, \boldsymbol{r}_2, \boldsymbol{r}_3$ производится по нижеследующим формулам. Радиус-вектора положений центров опорных первичных сфер $\boldsymbol{r}_1, \boldsymbol{r}_2, \boldsymbol{r}_3$ определяют основание пирамиды, а центр вторичного шара-зонда $\boldsymbol{r}$ ее вершину. Этот случий похож на случай первичной сферической перетяжки. Он даже проще, поскольку все ребра пирамиды, выходящие из ее вершины в текушем случае равны.

Радиус-вектор центра шара-зонда можно разложить по двум векторам - это радиус-вектор основания высоты пирамиды $\mathbf{p}_c$ и единичный вектор, перпендикулярный основанию пирамиды $\mathbf{z}$:

$$\mathbf{r} = \mathbf{p}_c \pm \mathbf{z} h \ . \qquad (29)$$

Эти два радиус-вектор определяются двумя формулами

$$\mathbf{z} = \frac{[(\mathbf{r}_3 - \mathbf{r}_1) \times (\mathbf{r}_2 - \mathbf{r}_1)]}{\|[(\mathbf{r}_2 - \mathbf{r}_1) \times (\mathbf{r}_3 - \mathbf{r}_1)]\|}\ , \qquad (30)$$

$$\mathbf{p}_c = \frac{(\mathbf{r}_1 + \mathbf{r}_2 + \mathbf{r}_3)}{3} + \frac{1}{6}\left[\left(\begin{array}{l} \frac{(\mathbf{r}_3 - \mathbf{r}_1)^2 - (\mathbf{r}_2 - \mathbf{r}_1)^2}{\|[(\mathbf{r}_3 - \mathbf{r}_1)\times(\mathbf{r}_2 - \mathbf{r}_1)]\|^2}[[(\mathbf{r}_3 - \mathbf{r}_1)\times(\mathbf{r}_2 - \mathbf{r}_1)]\times \mathbf{r}_1] + \\ + \frac{(\mathbf{r}_1 - \mathbf{r}_2)^2 - (\mathbf{r}_3 - \mathbf{r}_2)^2}{\|[(\mathbf{r}_1 - \mathbf{r}_2)\times(\mathbf{r}_3 - \mathbf{r}_2)]\|^2}[[(\mathbf{r}_1 - \mathbf{r}_2)\times(\mathbf{r}_3 - \mathbf{r}_2)]\times \mathbf{r}_2] + \\ + \frac{(\mathbf{r}_2 - \mathbf{r}_3)^2 - (\mathbf{r}_1 - \mathbf{r}_3)^2}{\|[(\mathbf{r}_2 - \mathbf{r}_3)\times(\mathbf{r}_1 - \mathbf{r}_3)]\|^2}[[(\mathbf{r}_2 - \mathbf{r}_3)\times(\mathbf{r}_1 - \mathbf{r}_3)]\times \mathbf{r}_3] \end{array}\right)\right], \qquad (31)$$

$$h^2 = (R_{pr} + R_{sec})^2 - (\mathbf{r}_1^2 + \mathbf{r}_2^2 + \mathbf{r}_3^2)/3 - \mathbf{p}_c^{\,2} + 2\mathbf{p}_c \cdot \frac{(\mathbf{r}_1 + \mathbf{r}_2 + \mathbf{r}_3)}{3}, \qquad (32)$$

где $R_{sec}$ - радиус сферы вторичной обкатки.

### 2.6.5 Определение массива координат центров и радиусов пар вторичных сферических фрагментов (сферы устойчивых положений), сглаживающих торы первичной обкатки при их самопересечении или наличии узких перешейков.

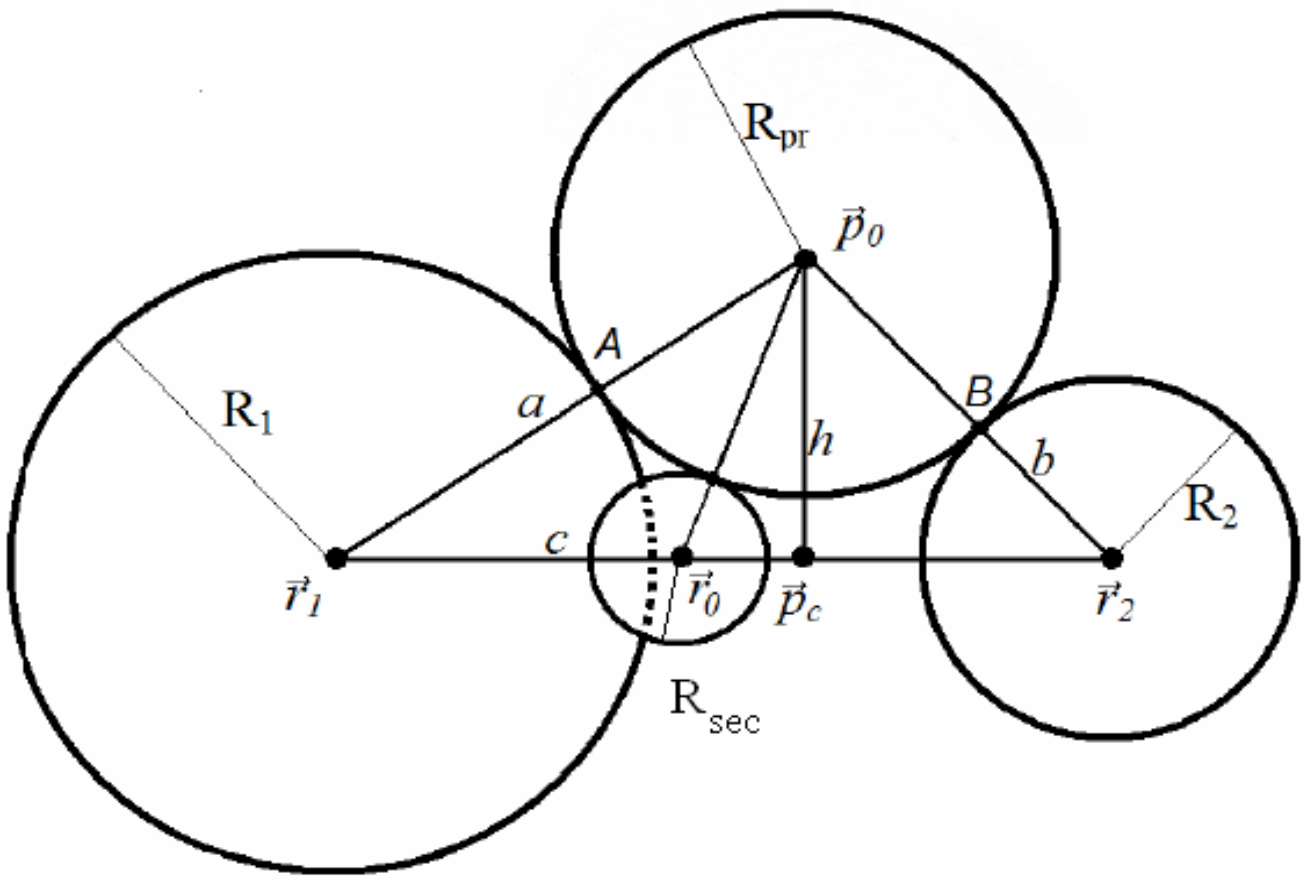

**Рис. 8** Вторичные сферы устойчивых положений. См. [41].

Для точек находящихся на поверхности сфер устойчивых положений (Рис. 8). Расстояние центра сферы до центра тора *d*:

$$d = \sqrt{(R_{pr} + R_{\sec})^2 - h^2}\ . \qquad (33)$$

Радиус-вектора $\boldsymbol{r}_0$ центров сфер устойчивых положений:

$$\boldsymbol{r}_0 = \boldsymbol{p}_c \pm d\boldsymbol{z}\ . \qquad (34)$$

### 2.6.6 Определение массива координат центров и радиусов вторичных сферических фрагментов (вторичная сферическая перетяжка) между тройками первичных сферических фрагментов .

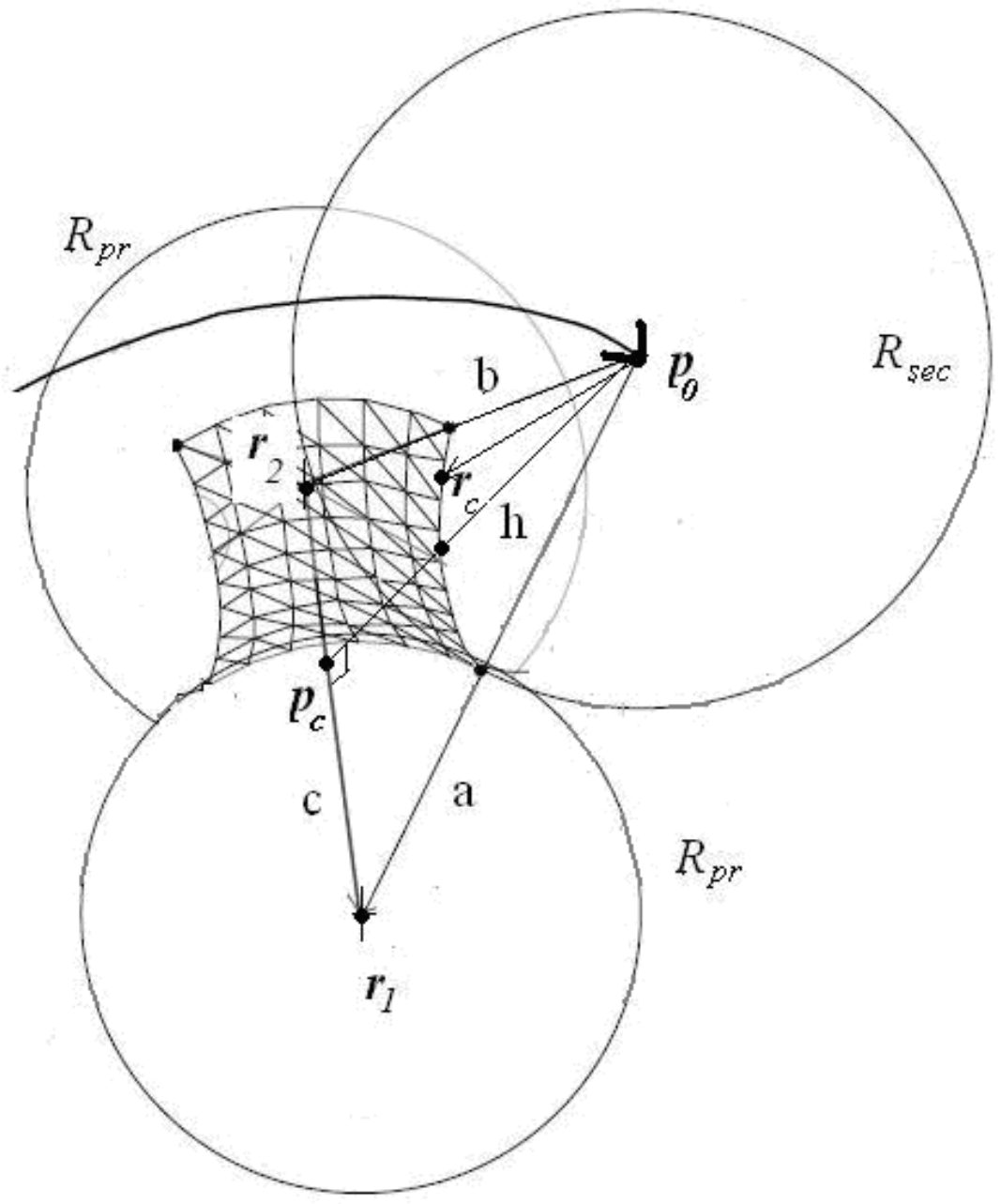

**Рис. 9.** Вторичная обкатка двух сфер первичной обкатки. Торы вторичной обкатки.

Тороидальные вторичные фрагменты (Рис. 9) поверхности образуются в процессе обкатки вторичной сферой обкатки двух первичных сферических сегментов. При этом центр вторичной сферы обкатки описывает круговую дугу или полную окружность в зависимости от расположения других первичных сфер или первичных торов. Тороидальный вторичный фрагмент заключён между точками касания вторичной сферы двух первичных сфер и положениями вторичной сферы обкатки, в которых она упирается в какой-либо другую (или даже несколько *одновременно*) первичную сферу или первичный тор.

Для описания вторичных тороидальных фрагментов используются следующие параметры.

Центр $\mathbf{p}_c$ и радиус $h$ окружности, вокруг которой описывает дугу центр сферы вторчной обкатки обкатки.

Локальный базис $x,y,z$, ось $z$ которого направлена параллельно прямой соединяющей центры опорных первичных сфер, а оси $x$ и $y$ определяют плоскость дуги центра вторичной сферы обкатки. При этом ось $x$ направлена либо на одну из сфер первичной обкатки, которая ограничивает обкатку вокруг данной пары атомов сфер первичной обкатки, либо (для ограничивающего движение первичного тора) на центр вторичной сферы, ограничивающий вращение.

Углы $\alpha$ и $\beta$ относительно оси $x$ начала и конца дуговой траектории центра сферы обкатки. Плоскость Рис. 9 задают центры сфер первичной обкатки $\boldsymbol{r}_1$ и $\boldsymbol{r}_2$ и точка поверхности $\boldsymbol{r}_s$: $R_{sec}$-радиус сферы вторичной обкатки, $R_{pr}$– радиус сферы первичной обкатки.

Стороны треугольника, образуемого центрами сфер и их производные:

$$a = b = R_{pr} + R_{sec} \, ; \; c = \left|\boldsymbol{r}_2 - \boldsymbol{r}_1\right| . \qquad (35)$$

Высота треугольника или, что тоже, радиус круговой траектории центра сферы вторичной обкатки:

$$\mathrm{h} = \sqrt{\mathrm{a}^2 - \frac{\mathrm{c}^2}{4}} \; . \qquad (36)$$

Центр круговой траектории центра сферы вторичной обкатки или, что тоже, радиус-вектор точки основания высоты треугольника $\boldsymbol{p}_c$:

$$\boldsymbol{p}_c = \frac{\boldsymbol{r}_1 + \boldsymbol{r}_2}{2} . \qquad (37)$$

$\boldsymbol{x}, \boldsymbol{y}, \boldsymbol{z}$ - локальный базис парной перетяжки (центр координат в основании высоты $\boldsymbol{p}_c$, ось z в направлении от первой опорной первичной сферы ко второй):

$$\boldsymbol{z} = \frac{\boldsymbol{r}_2 - \boldsymbol{r}_1}{c}, \; (\boldsymbol{x} \cdot \boldsymbol{z}) = 0, \; (\boldsymbol{y} \cdot \boldsymbol{z}) = 0, \; (\boldsymbol{x} \cdot \boldsymbol{y}) = 0 \; . \qquad (38)$$

Если первичная сфера с координатами $\vec{r}_3$ ограничивает обкатку вокруг первичных сфер $r_1$ и $r_2$, то определение базисных векторов задано выражениями:

$$\boldsymbol{l} = \frac{\boldsymbol{r}_3 - \boldsymbol{p}_c}{|\boldsymbol{r}_3 - \boldsymbol{p}_c|}, \quad \boldsymbol{y} = [\boldsymbol{z} \times \boldsymbol{l}], \quad \boldsymbol{x} = [\boldsymbol{z} \times \boldsymbol{y}]. \qquad (39)$$

Если вторичной сфера ограничивает обкатку вокруг первичных сфер $r_1$ и $r_2$ выбрана для задания направления оси x, то определение базисных векторов задано выражениями:

$$\boldsymbol{x} = \frac{\boldsymbol{p}_{0,\alpha} - \boldsymbol{p}_c}{|\boldsymbol{p}_{0,\alpha} - \boldsymbol{p}_c|}, \quad \boldsymbol{y} = [\boldsymbol{z} \times \boldsymbol{x}], \qquad (40)$$

где $\boldsymbol{p}_{0,\alpha}$ - радиус-вектор центра ограничивающей вторичной сферы.

Расчёт углов $\alpha$ и $\beta$ производится по формулам:

$$\alpha = arccos(\boldsymbol{x} \cdot \frac{\boldsymbol{p}_{0,\alpha} - \boldsymbol{p}_c}{|\boldsymbol{p}_{0,\alpha} - \boldsymbol{p}_c|}), \quad \beta = arccos(\boldsymbol{x} \cdot \frac{\boldsymbol{p}_{0,\beta} - \boldsymbol{p}_c}{|\boldsymbol{p}_{0,\beta} - \boldsymbol{p}_c|}), \qquad (41)$$

где $\boldsymbol{p}_{0,\alpha}$ и $\boldsymbol{p}_{0,\beta}$ - координаты центра вторичной сферы обкати при касании трёх первичных сфер на начале и конце дуги обкатки вокруг пары первичных сфер или координаты центров вторичных сфер устойчивого положения. Проекция нормали на локальный базис для торов вторичной обкатки (нормаль направлена внутрь тора):

$$\begin{aligned} &\boldsymbol{n}_s \cdot \boldsymbol{z} = -\gamma_s, \\ &(\boldsymbol{n}_s \cdot \boldsymbol{x}) = ((-\sqrt{\boldsymbol{n}_{sx}^2 + \boldsymbol{n}_{sy}^2}\, \boldsymbol{e}_s) \cdot \boldsymbol{x}) = -\alpha_s \sqrt{1 - \gamma_s^2}, \\ &(\boldsymbol{n}_s \cdot \boldsymbol{y}) = ((-\sqrt{\boldsymbol{n}_{sx}^2 + \boldsymbol{n}_{sy}^2}\, \boldsymbol{e}_s) \cdot \boldsymbol{y}) = -\beta_s \sqrt{1 - \gamma_s^2}. \end{aligned} \qquad (42)$$

$$\boldsymbol{n}_s = -\alpha_s \cdot \sqrt{1 - \gamma_s^2} \cdot \boldsymbol{x} - \beta_s \cdot \sqrt{1 - \gamma_s^2} \cdot \boldsymbol{y} - \gamma_s \cdot \boldsymbol{z}. \qquad (43)$$

## 2.7 Поверхность SAS

3 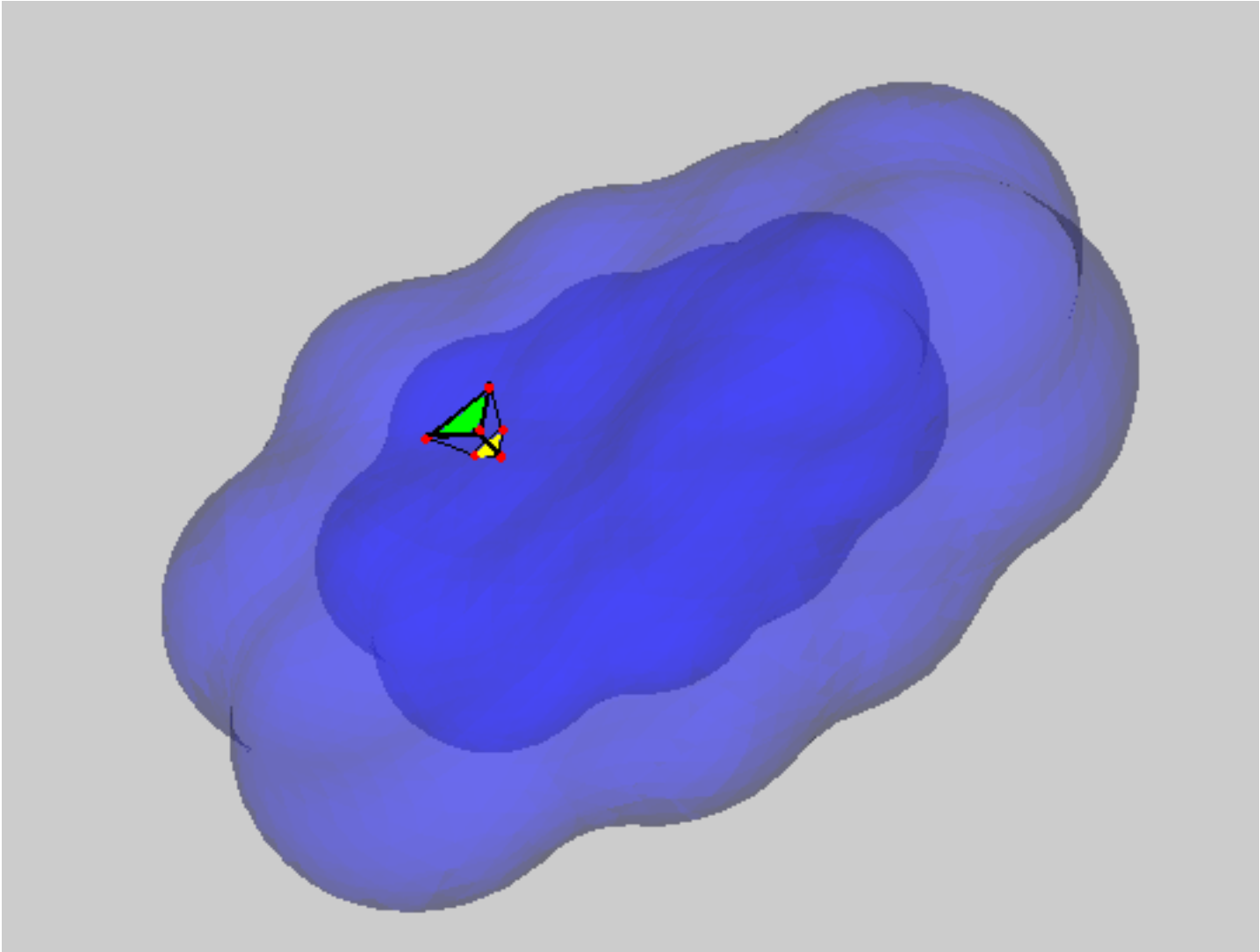

**Рис. 10**. Преобразование поверхности типа SES в поверхность типа SAS

Пусть имеется поверхность после только *первичной* обкатки. Тогда поверхность SAS (Рис. 10) получается из нее переносом всех точек поверхности вдоль внешней нормали на расстояние, равное радиусу первичной обкатки ($R_{pr}$). При этом сегменты сфер атомов с радиусом $R_{atom}$ преобразуются в сегменты сфер SAS с радиусом $R_{atom}$+ $R_{pr}$, торы первичной обкатки выраждаются в линии пересчения сфер, сферические сегметы тройных точек вырождаются в точки пресечения трех (или более при вырождении) сфер.

.

## 3. Выводы.

Представленный алгоритм и основанные на нем программы позволяют быстро и надежно строить максимально гладкую поверхность вокруг любых молекул, включая белки, а затем делать ее триангуляцию. Поверхность получается раскрашенной в зависимости от типа ближайших атомов, что удобно для визуализации, и триангулированной, что позволяет не только вычислять ее поверхность и ограниченный ею объем, но и решать с хорошей точностью интегральные уравнения, используемые в континуальной модели растворителя. Полученная с помощью данной программы поверхность может применяться как для целей визуализации молекулы, это особенно актуально для больших белковых молекул, так и для целей вычисления сольватационных вкладов в энергию взаимодействия молекул друг с другом при наличии внешней среды.

**Благодарности.**

Данная работа является более детальным изложением и дальнейшим развитием статьи [41]. Авторы выражают глубокую признательность всем авторам этой публикации за проделанную работу, которая послужила основой и для данной статьи.

Список литературы